\documentclass[9pt,twocolumn]{extarticle}
\pdfoutput=1
\usepackage{soul}
\usepackage{titlesec}
\usepackage{cite}
\usepackage{float}
\usepackage{dsfont}
\usepackage{graphicx}
\usepackage{subcaption}
\usepackage[margin=0.9in]{geometry}
\usepackage[usenames,dvipsnames]{color}
\usepackage[colorlinks,linkcolor=Blue,urlcolor=Blue,citecolor=Blue]{hyperref}
\usepackage{amsmath,amssymb}
\usepackage[squaren]{SIunits}

\usepackage{mathpazo}
\usepackage{courier}
\normalfont
\usepackage[T1]{fontenc}
\usepackage{caption}
\usepackage{upgreek}

\newcommand{\ad}[1]{\textsuperscript{#1}\kern-2pt}

\makeatletter
\def\blx@maxline{77}
\makeatother

\usepackage{capt-of}

\def\mytitle{Diverse polymorphs and phase transitions in van der Waals In$_2$Se$_3$}

\title{\vspace{-1.0cm}\Huge\textbf{\textrm{\mytitle}}}

\author{Mingfeng Liu,$^{1,\star}$ Jiantao Wang,$^{1,2,\star}$  Peitao Liu,$^{1,\star,\dagger}$ Qiang Wang,$^{1,2}$ Zhibo Liu,$^{1}$  Yan Sun,$^{1}$ and Xing-Qiu Chen$^{1,\dagger}$ }

\begin{document}
	\twocolumn[{
		\maketitle
		\vspace{-5mm}
		\begin{center}
			\begin{minipage}{1\textwidth}
				\begin{center}
					\textit{
						\textsuperscript{1} Shenyang National Laboratory for Materials Science, Institute of Metal Research, Chinese Academy of Sciences, 110016 Shenyang, China
						\\\textsuperscript{2} School of Materials Science and Engineering, University of Science and Technology of China, Shenyang 110016, China
						\vspace{5mm}
						\\{$\star$} These authors contribute equally.
						\\{$\dagger$} Corresponding to: ptliu@imr.ac.cn, xingqiu.chen@imr.ac.cn
						\vspace{5mm}
					}
				\end{center}
			\end{minipage}
		\end{center}

		\setlength\parindent{13pt}
		\begin{quotation}
			\noindent
			\section*{Abstract}
Van der Waals In$_2$Se$_3$ has garnered significant attention due to its unique properties and wide applications
associated with its rich polymorphs and polymorphic phase transitions.
Despite extensive studies, the vast complex polymorphic phase space remains largely unexplored, and
the underlying microscopic mechanism for their phase transformations remains elusive.
Here, we develop a highly accurate, efficient, and reliable machine-learning potential (MLP),
which not only facilitates accurate exploration of the intricate potential energy surface (PES),
but also enables us to conduct large-scale molecular dynamics (MD) simulations with first-principles accuracy.
We identify the accurate structure of the $\beta''$ polymorph and uncover
several previously unreported $\beta'$ polymorph variants exhibiting dynamic stability and competing energies,
which are elucidated by characteristic flat imaginary phonon bands and the distinctive Mexican-hat-like PES in the $\beta$ polymorph.
Through the MLP-accelerated MD simulations,
we directly observe the polymorphic phase transformations among the $\alpha$, $\beta$, $\beta'$, and $\beta''$ polymorphs
under varying temperature and pressure conditions,
and build for the first time an \emph{ab initio} temperature-pressure phase diagram,
showing good agreement with experiments.
Furthermore, our MD simulations reveal a novel strain-induced reversible phase transition
between the $\beta'$ and $\beta''$ polymorphs.
This work not only unveils diverse polymorphs in van der Waals In$_2$Se$_3$,
but also provides crucial atomic insights into their phase transitions,
opening new avenues for the design of novel functional electronic devices.
\end{quotation} 	
}] 	 	 	
\newpage 	
\clearpage

\section*{Introduction}

Indium selenide (In$_2$Se$_3$) has prompted significant interest because of
its unique ferroic characteristics~\cite{DingNC2017,NanoLett_2017_Zhou,PRLXiao2018,NanoLett_2018_Cui,NanoLetters_Poh2018,AFM_2018,
NanoResearchIo2020, MaterHoriz2021, ShiLiu2024_ACSNano,SA_Zheng2018,PRL_AFE2020,NC_ferroelasticity_Xu2021,NatureReviewsMaterials_Zhang2023, JPCL_AFE_Wu2023,SA_AFE_2024,ACSNano_Wang2024,AM_Review_Li2024},
rich polymorphs with distinct polytypes~\cite{Sreekumar2006,Kupers2018,ChemistryMaterials_Liu2019,
ACSNano_Review_Li2021,InfoMat_Review_XianbinLi2022,APR_review_Han2024,ChemicalReviews_Tan2023},
and complex polymorphic phase transitions~\cite{ACSNano_Review_Li2021,InfoMat_Review_XianbinLi2022,
ChemicalReviews_Tan2023,APR_review_Han2024,Nanophotonics2024}
under various external stimuli,
such as temperature~\cite{van1975,Manolikas1988,JipingYe_1998,NanoLetters_Tao2013,ACSNano_Zhang2019,ACS_Applied_Lyu2022,NanoLetters_Wu2023}, laser~\cite{Igo2019,AM_LaserLi2022,Wan2022,Guo2025},
electric field~\cite{AdvancedScience_Chen2021,AM_electric_Zhang2022,NC_beta1_beta2_electric_field_Zhang2024,NC_JiWei_Wu2024,NM_Zhang2025}, pressure~\cite{APL_pressure_Rasmussen2013,APL_pressure_Ke2014,Zhao2014_pressure,InorganicChemistry_Pressure_Vilaplana2018,NC_Ppressure_Tang2023}, strain~\cite{APL_strain_Dong2020,ScienceAdvances_Zheng2022,NatureNanotechnology_Han2023},
and Se vacancy~\cite{NatureNanotechnology_Han2023}
as well as thickness variations~\cite{ScienceAdvances_Zheng2022,NanoLettersChen2023}.
The correlated ferroicity and polymorphism in van der Waals (vdWs) In$_2$Se$_3$
therefore provide vast opportunities in tailoring its properties for a wide range of applications~\cite{ACSNano_Review_Li2021,InfoMat_Review_XianbinLi2022,APR_review_Han2024},
including
ferroelectrics~\cite{DingNC2017,NanoLett_2017_Zhou,PRLXiao2018,NanoLett_2018_Cui,NanoLetters_Poh2018,AFM_2018,SA_Zheng2018,
NanoResearchIo2020,MaterHoriz2021,ShiLiu2024_ACSNano},
ferroelasticity~\cite{NC_ferroelasticity_Xu2021},
thermoelectrics~\cite{smallreview_2014,WangJPCL2017},
photodetectors~\cite{Lin2013_JACS,doi:10.1021/nn9012466,10.1063/1.3669513,AFM_Xue2020,AFM_Yang2022},
neuromorphic devices~\cite{NC_Wang2021,NatureElectronics_Liu2022},
solar cells~\cite{Peng2007,doi:10.1021/acs.jpclett.2c02975},
and phase-change memories~\cite{LEE2005196,Yu2007,Huang2014,AM_Choi2017,Nanophotonics2024}.

Among the various polymorphs of In$_2$Se$_3$, the $\alpha$ polymorph has mostly been studied
due to its unique property of thickness-scalable room-temperature in-plane and out-of-plane ferroelectricity~\cite{DingNC2017,NanoLett_2017_Zhou,PRLXiao2018,NanoLett_2018_Cui,NanoLetters_Poh2018,AFM_2018,NanoResearchIo2020,MaterHoriz2021,ShiLiu2024_ACSNano}.
The $\alpha$-In$_2$Se$_3$ processes two different polytypes, i.e., hexagonal (2H) and rhombohedral (3R) structures,
which differ in the stacking sequences of Se-In-Se-In-Se quintuple layer blocks
separated by vdWs gaps~\cite{ChemistryMaterials_Liu2019,ChemicalReviews_Tan2023}.
In addition to the $\alpha$ polymorph, the $\beta$ polymorph and its two variants $\beta'$ and $\beta''$ have also attracted growing interest.
The $\beta$-In$_2$Se$_3$ manifests three distinct polytypes: trigonal (1T), 2H, and 3R structures.
It can be derived from the $\alpha$ polymorph by heating to temperatures above 473 K~\cite{van1975,Manolikas1988,JipingYe_1998,NanoLetters_Tao2013,ACS_Applied_Lyu2022}
or by applying a pressure up to 0.7$\sim$0.8 GPa~\cite{APL_pressure_Rasmussen2013,APL_pressure_Ke2014}.

The $\beta'$-In$_2$Se$_3$ polymorph is of particular interest because of its distinctive superstructure
characterized by unusual nanostripe orderings~\cite{SA_Zheng2018,PRL_AFE2020,NC_ferroelasticity_Xu2021,
NatureReviewsMaterials_Zhang2023, JPCL_AFE_Wu2023,SA_AFE_2024,ACSNano_Wang2024}.
It can be acquired by cooling the high-temperature $\beta$ polymorph below 473 K,
and such thermally-driven transformation is reversible~\cite{van1975,Manolikas1988,SA_Zheng2018,PRL_AFE2020,NC_ferroelasticity_Xu2021}.
Xu~\emph{et al.} have identified the nanostripe-ordered $\beta'$-In$_2$Se$_3$
as a two-dimensional antiferroelectric material, with antiparallel displacements of central Se atoms between adjacent nanostripes~\cite{PRL_AFE2020}.
The in-plane antiferroelectric distortion of $\beta'$-In$_2$Se$_3$ further results in ferroelasticity and the formation of domain walls (DWs)~\cite{NC_ferroelasticity_Xu2021}.
Wang~\emph{et al.} have extended the findings of Xu~\emph{et al.}
by showing that that both zero or nonzero net polarization can be obtained,
depending on whether the opposite in-plane ferrielectric orderings within neighboring nanostripes are compensated or noncompensated~\cite{ACSNano_Wang2024}.

As an another variant of the $\beta$ polymorph, $\beta''$-In$_2$Se$_3$ is derived by cooling the $\beta'$ polymorph below 180 K~\cite{ACSNano_Zhang2019}.
Interestingly, the $\beta'$ and $\beta''$ polymorphs can be reversibly interconverted by tuning temperature~\cite{ACSNano_Zhang2019}
or by applying an electric field~\cite{AM_electric_Zhang2022,NC_beta1_beta2_electric_field_Zhang2024}.
The $\beta''$-In$_2$Se$_3$ exhibits a typical zig-zag striped morphology,
which is driven by a 2$\times\sqrt{3}$ surface reconstruction~\cite{ACSNano_Zhang2019,AdvancedScience_Chen2021,AM_electric_Zhang2022,NC_beta1_beta2_electric_field_Zhang2024}.
Recently, a novel chiral Star-of-David charge density wave (CDW) phase with energy comparable to that of the $\beta''$ polymorph
has been predicted in the vdWs In$_2$Se$_3$ monolayer~\cite{NC_cal_Huang2024}.

Despite extensive experimental and theoretical studies on the polymorphs and polymorphic phase transitions in vdWs In$_2$Se$_3$,
the vast polymorphic phase space remains largely unexplored.
This is attributed to the existence of multiple polytypes and
the presence of prominent flat imaginary phonon frequencies in the optical branch of the $\beta$ polymorph at 0 K,
leading to a Mexican-hat-like potential energy surface (PES) for the middle-layer Se atoms~\cite{APR_cal_Huang2021,NC_cal_Huang2024}.
While first-principles calculations based structure search can be employed to explore the polymorphic phase space,
the substantial computational costs involved make this approach prohibitive, especially for bulk In$_2$Se$_3$ typically consisting of large unit cells.
Concerning the polymorphic phase transitions, current atomic insights mostly focus on the structural transition
between the $\alpha$ and $\beta$ monolayers~\cite{2DMaterials_Liu2019,Nanoscale_Soleimani2020,LiuShi_PRB2021,APR_cal_Huang2021},
whereas the underlying microscopic phase transition mechanism
among the $\beta$ polymorph and its two variants $\beta'$ and $\beta''$ remains elusive,
primarily due to inadequate spatial and temporal resolutions in experiments
and limited time and length scales accessible for \emph{ab initio} molecular dynamics (MD) simulations.
In particular, the formation and movement of DWs at the atomic scale have not been fully understood,
hindering their application in nanoelectronics.
Furthermore, a thermodynamic temperature-pressure phase diagram for In$_2$Se$_3$ is currently still lacking,
which is essential for understanding its stability, phase transitions, and potential technological applications.

Aiming at addressing these challenges, we have developed a highly accurate, efficient, and reliable
moment tensor machine-learning potential (MLP) for bulk vdWs In$_2$Se$_3$
through active learning and optimization of basis sets.
The MLP allows us to effectively explore the complex PES of In$_2$Se$_3$,
resulting in the identification of accurate structure of the $\beta''$ polymorph
and several previously unreported $\beta'$ polymorph variants across all 1T, 2H, and 3R polytypes.
These newly discovered polymorphs exhibit dynamic stability and possess competing energies,
which can be understood through the presence of flat imaginary phonon modes and the Mexican-hat-like PES observed in the $\beta$ polymorph.
Additionally, the developed MLP enables us to conduct large-scale MD simulations with \emph{ab initio} accuracy,
allowing for direct observation of the phase transitions from $\alpha$ to $\beta$, $\beta$ to $\beta'$, and $\beta'$ to $\beta''$
under varying temperature and pressure conditions.
These atomistic simulations provide crucial microscopic insights into the dynamical phase transformations,
including nucleation and growth, as well as the formation and movement of the DWs.
Moreover, we have for the first time built an \emph{ab initio} temperature-pressure phase diagram
using thermodynamic integration method, aligning well with experimental findings.
Furthermore, our MD simulations reveal the strain-induced reversible phase transformations between the $\beta'$ and $\beta''$ polymorphs,
offering a new avenue for polarization modulations.
By exploring the diverse polymorphs and phase transitions in vdWs In$_2$Se$_3$, this work pays the way for the design of novel functional electronic devices.

\begin{figure*}[ht!]
	\centering
\includegraphics[width=0.8\textwidth]{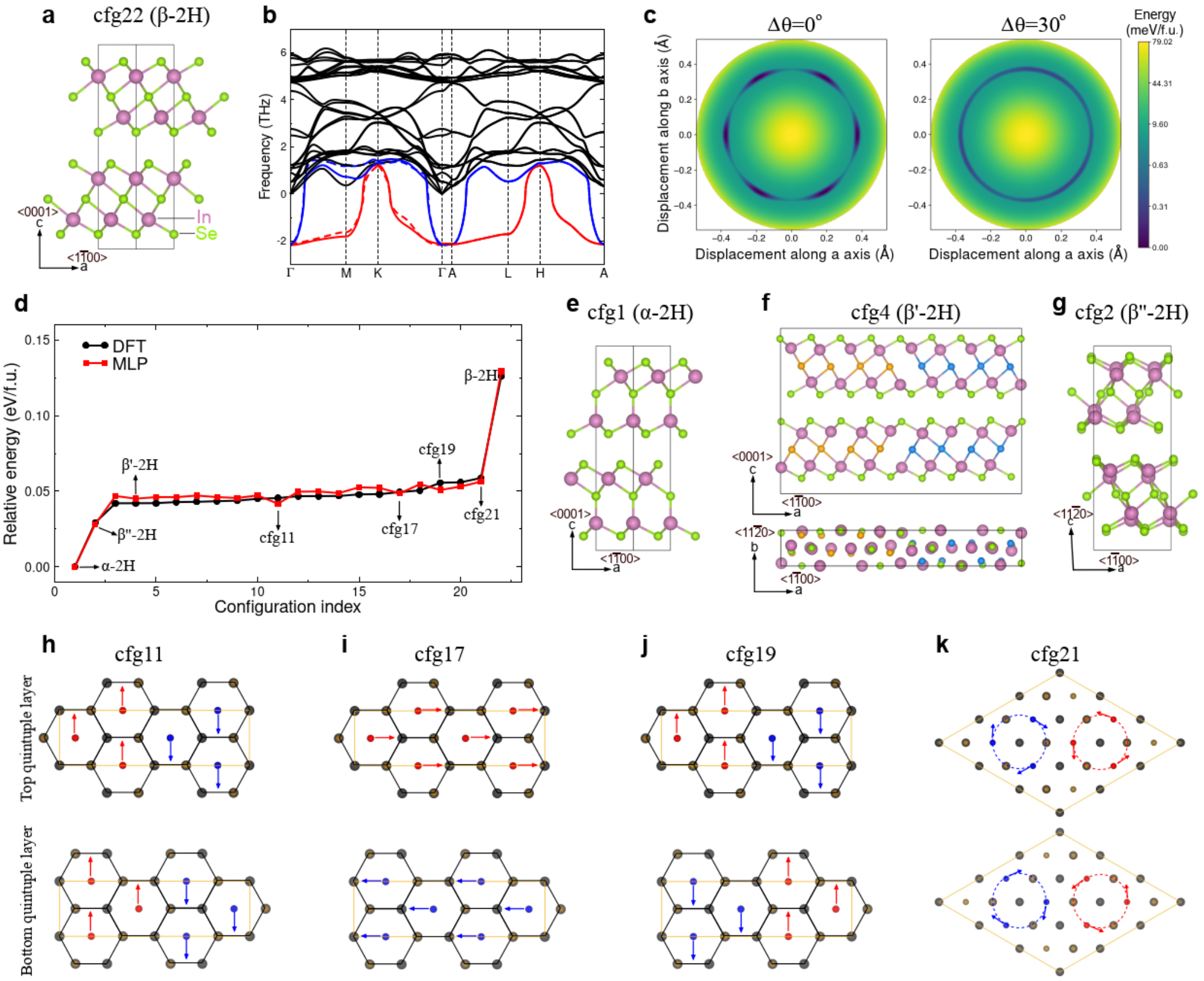}
		\caption{\textbf{| Diverse polymorphs of vdWs In$_2$Se$_3$.}
			\textbf{a} Crystal structure of the ideal $\beta$-2H phase.
			\textbf{b} Phonon dispersions of $\beta$-2H at 0 K.
Two pairs of optical phonon branches with imaginary frequencies are highlighted by the solid (red) and dashed (blue) lines.
The solid and dashed curves represent in-phase and anti-phase polarizations between the two quintuple layers, respectively.
			\textbf{c} The Mexican-hat-like PES when adjusting the positions of the central Se atoms in both quintuple layers of $\beta$-2H
for two phase difference angles: $\phi$=$0\degree$ and $\phi$=$30\degree$.
            \textbf{d} Relative energies of identified polymorphs in the 2H polytype with respect to the $\alpha$-2H phase predicted by both DFT and MLP.
            The polymorphs are arranged in ascending order of DFT-calculated energies.
            The crystal structures of all identified polymorphs with the 2H polytype are presented in Supplementary Fig.~2.
            \textbf{e}, \textbf{f}, \textbf{g} Crystal structures of $\alpha$-2H, the 4$d$-4$\bar{d}$/4$d$-4$\bar{d}$ variant of $\beta'$-2H, and $\beta''$-2H.
            Note that the central Se atoms exhibiting antiparallel displacements along the $\langle11\bar{2}0\rangle$ direction are highlighted in orange and blue.
            \textbf{h}-\textbf{k} Schematic structures of selected polymorphs in the 2H polytype.
            The arrows denote the displacements of the central Se atoms indicated by the small red and blue circles.
            The large circles represent In atoms, while the small circles denote Se atoms.
            The orange small circles denote the top Se atoms, and the orange lines indicate the unit cell.
	}
	\label{Figure1}
\end{figure*}

\section*{Results}
\noindent\textbf{Development of MLP } \\

To establish a reliable MLP that can accurately capture the complex PES and various polymorphic phase transitions of bulk vdWs In$_2$Se$_3$,
it is crucial to create a representative training dataset that encompasses the broad phase space of interest.
To achieve this, we followed the scheme outlined in our previous works~\cite{PhysRevLett.130.078001,Liu2024,caoyu2024}.
Specifically, the training structures were initially sampled using an
on-the-fly active learning scheme based on the Bayesian linear regression~\cite{JinnouchiPRL2019,JinnouchiPRB2019}.
This enabled us to efficiently sample the phase space and automatically collect the representative training structures
during the \emph{ab initio} MD simulations under varying temperature (0$\sim$900 K) and pressure (0$\sim$6 GPa) conditions.
The starting structures were taken from bulk $\alpha$, $\beta$, $\beta'$, and $\beta''$ phases with different polytypes (i.e., 1T, 2H, and 3R).
Then, the training dataset was iteratively expanded using the active learning method~\cite{Alexander2016,Podryabinkin2017}
as implemented in the MLIP package~\cite{Novikov_2021}.
Those unseen structures with large extrapolation grade exceeding 10
were extracted from extensive MD simulations and the relaxation process of perturbed structures~\cite{Novikov_2021}.
The energies, forces, and stresses of the sampled structures were computed from first-principles density functional theory (DFT) calculations, as detailed in the "Methods" section.
Note that although the present work primarily focus on the phase transformations among the $\alpha$, $\beta$, $\beta'$, and $\beta''$ polymorphs of In$_2$Se$_3$,
we also incorporated additional 2,133 InSe structures, including various phases such as sphalerite, wurtzite, $\beta$, and $\gamma$, without and with In vacancies.
These pristine and defective InSe structures share some local structural similarities with In$_2$Se$_3$,
but they also introduce new local atomic motifs, thereby enhancing the transferability of the MLP.
Furthermore, the intralayer split structures reported in Refs.~\cite{NC_JiWei_Wu2024,NM_Zhang2025} were also included in the training dataset.
The eventual training dataset contained 5,262 structures and covered a wide range of phase space,
as indicated by the kernel principal component analysis~\cite{doi:10.1021/acs.accounts.0c00403} (Supplementary Fig.~1a).
To enhance the efficiency, a moment tensor potential was fitted with the optimized basis sets
by refining the contraction process of moment tensors using our in-house code (IMR-MLP)~\cite{wang2024}.
For more details on computational settings, we refer to the "Methods" section.

The developed MLP was validated on a test dataset containing 628 structures.
These tested structures including $\alpha$, $\beta$, $\beta'$ and $\beta''$ phases with different stacking sequences
were randomly sampled from extensive MD simulations at varied temperatures and pressures,
and covered only a fraction of the phase space represented by the training dataset (Supplementary Fig.~1a).
The validation (training) root-mean-square-errors (RMSEs) on energies, forces,
and stress tensors are 0.61 (1.17) meV/atom, 0.039 (0.067) eV/\AA, and 0.31 (0.68) kbar, respectively, showing high accuracy of the MLP (Supplementary Fig.~1).
Furthermore, the accuracy of the MLP was rigorously validated by predicting the lattice parameters,
energy differences, energy-volume curves, and phonon dispersions of various polymorphs,
exhibiting excellent agreement with the underlying DFT results as well as the experimental data, as will be discussed in detail in the following sections.

\vspace{3mm}
\noindent\textbf{Exploration of diverse polymorphs }\\
We start our exploration of the diverse polymorphs of vdWs In$_2$Se$_3$ by analyzing the phonon dispersions of the $\beta$ phase.
Without loss of generality, our discussion will focus specifically on the 2H polytype unless explicitly noted otherwise.
The $\beta$-2H polymorph is a high-temperature phase with a centrosymmetric structure,
where all the In atoms are octahedrally coordinated with adjacent Se atoms (Fig.~\ref{Figure1}a).
Expectedly, the $\beta$-2H phase at 0 K displays dynamic instability, as evidenced by the presence of large flat imaginary frequencies
in two pairs of optical phonon branches (Fig.~\ref{Figure1}b).
One pair, shown in blue, corresponds to the polar distortion along the $\langle 1\bar{1}00 \rangle$ direction,
while the other pair, depicted in red, corresponds to the polar distortion along the $\langle 11\bar{2}0 \rangle$ direction.
This observation implies that multiple polymorphs can be formed by modulating the $\beta$-2H structure
according to the wavevectors of the imaginary phonon modes~\cite{Venkataraman1979,Pallikara2022}.
Indeed, the ferroelectric $\alpha$-2H phase with two distinct coordinated In atoms (Fig.~\ref{Figure1}e)
can be obtained by stabilizing the soft mode at $\Gamma$.
Similarly, by freezing the soft modes along the $\Gamma$-$M$ direction, various variants of the $\beta'$ polymorph can emerge.
Notably, stabilizing the soft modes at ($x$, 0, 0) with specific fractional values of $x$ such as 1/2, 2/5, 1/3, 2/7, 1/4, 2/9, and 1/5
results in the generation of distinct variants of $\beta'$, denoted as
2$d$-2$\bar{d}$/2$d$-2$\bar{d}$,
2$d$-3$\bar{d}$/2$d$-3$\bar{d}$,
3$d$-3$\bar{d}$/3$d$-3$\bar{d}$,
3$d$-4$\bar{d}$/3$d$-4$\bar{d}$,
4$d$-4$\bar{d}$/4$d$-4$\bar{d}$,
4$d$-5$\bar{d}$/4$d$-5$\bar{d}$,
and 5$d$-5$\bar{d}$/5$d$-5$\bar{d}$, respectively.
Here, the notation $md$-$n\bar{d}$/$md$-$n\bar{d}$ represents the two adjacent nanostripes with widths of $md_{1\bar{1}00}$ and $nd_{1\bar{1}00}$,
featuring antiparallel displacements of central Se atoms along the $\langle11\bar{2}0\rangle$ direction.
As an example, the crystal structure of the 4$d$-4$\bar{d}$/4$d$-4$\bar{d}$ superstructure is illustrated in Fig.~\ref{Figure1}f.

We note that, with the exception of the 3$d$-4$\bar{d}$/3$d$-4$\bar{d}$ superstructure,
all other superstructures introduced above have been experimentally identified~\cite{van1975,Lin2013_JACS,PRL_AFE2020,NC_ferroelasticity_Xu2021,ACSNano_Wang2024}.
These identified variants of the $\beta'$-2H phase exhibit competing energies (Fig.~\ref{Figure1}d),
and their predicted lattice parameters and periodic lengths agree well with the experimental data (Supplementary Table~1, Supplementary Table~2).
The existence of these superstructures with competing energies allows for the possibility of achieving local zero or nonzero net polarization within a sample,
depending on whether the antiparallel in-plane ferroelectric arrangements in adjacent nanostripes are compensated or noncompensated.
This may reconcile the conflicting experimental results concerning the ferroelectric or antiferroelectric nature of $\beta'$~\cite{SA_Zheng2018,ACSNano_Zhang2019,PRL_AFE2020}.

In fact, the superstructures identified are merely special manifestations of the complex PES of In$_2$Se$_3$.
Huang~\emph{et al.} have demonstrated the presence of a Mexican-hat-like PES for the central-layer Se atoms in the In$_2$Se$_3$ monolayer~\cite{APR_cal_Huang2021},
and predicted a novel chiral Star-of-David CDW phase~\cite{NC_cal_Huang2024}.
Here, we expand upon Huang's findings to encompass a broader scope by considering the stacking sequences of the blocks,
providing a vast array of tunable degrees of freedom.
Fig.~\ref{Figure1}c illustrates the PES when adjusting the positions of the central Se atoms in both quintuple layers of $\beta$-2H,
considering two specific phase difference angles $\phi$=$0\degree$ and $\phi$=$30\degree$.
The PESs for the additional symmetry-inequivalent phase difference angles $\phi$=$60\degree$ and $\phi$=$90\degree$ exhibit analogous patterns (Supplementary Fig.~3).
Evidently, in all cases, the PES displays a distinct Mexican-hat-like profile, featuring multiple minima, suggesting the presence of metastable phases.
Indeed, we verified the dynamical stability of all identified phases through phonon calculations (Supplementary Fig.~4).
The schematic structures of some of these metastable phases are shown in Fig.~\ref{Figure1}h-k,
while all identified structures in the 2H polytype are provided in Supplementary Fig.~2.
In addition to the ferroelectric phase, various intriguing variants of the antiferroelectric phase can be observed.
These include the A-type antiferroelectric configuration (Fig.~\ref{Figure1}i), characterized by intralayer ferroelectricity and interlayer antiferroelectricity,
as well as the G-type antiferroelectric configuration (Fig.~\ref{Figure1}j), which exhibits a checkboard-like antiferroelectric arrangement.
The Star-of-David CDW phase predicted in the In$_2$Se$_3$ monolayer~\cite{NC_cal_Huang2024}
has also been reproduced in the bulk 1T polytype (Supplementary Fig.~5n), which is demonstrated to be dynamically stable (Supplementary Fig.~6n).
Interestingly, a novel metastable phase with chirally-ordered central Se atoms is predicted (see Fig.~\ref{Figure1}k).

\begin{figure}
	\centering
\includegraphics[width=0.46\textwidth]{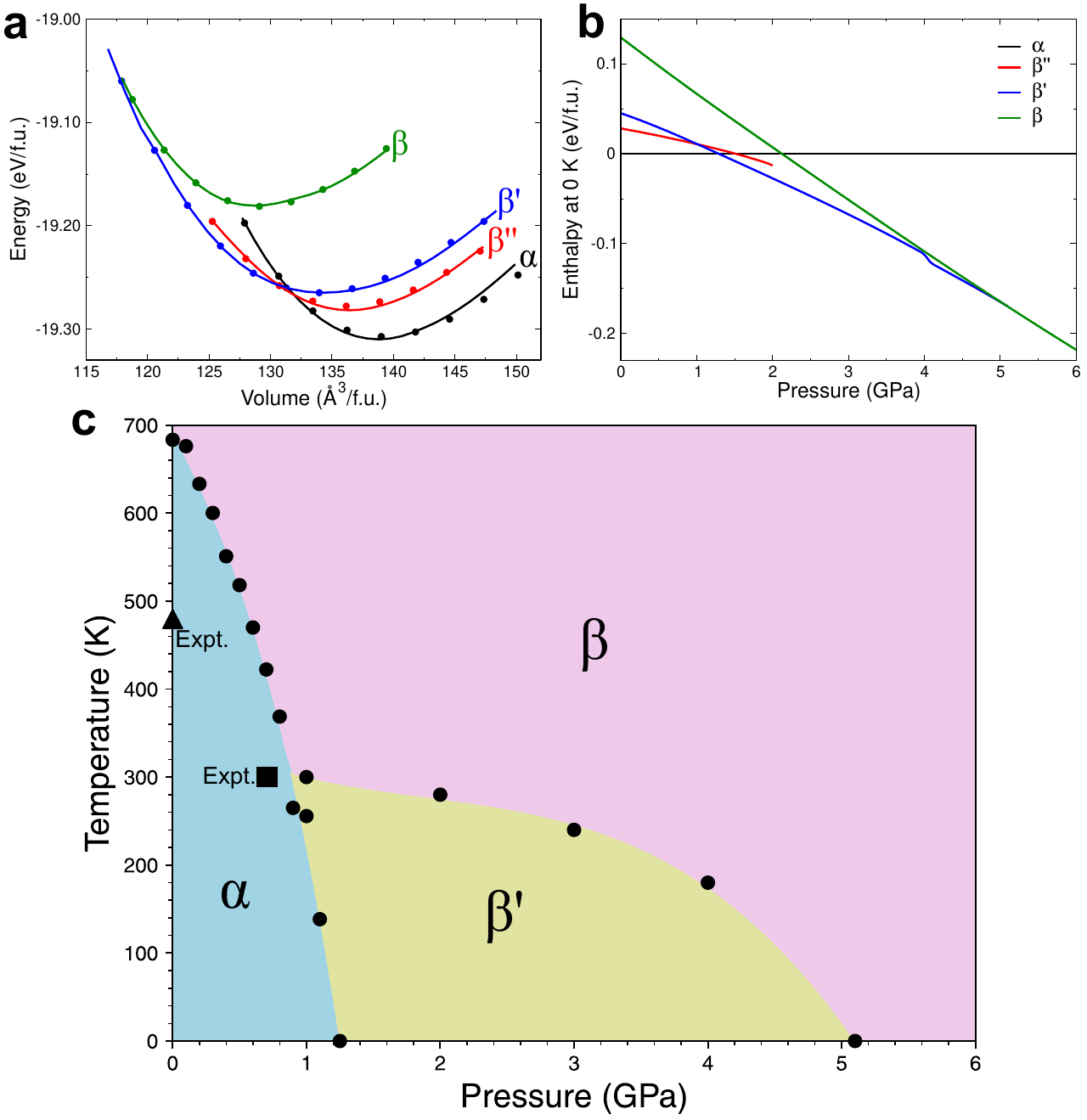}
		\caption{\textbf{| Thermodynamics of In$_2$Se$_3$.}
			\textbf{a} Predicted energy-volume curves at 0K for $\alpha$, $\beta$, the 4$d$-4$\bar{d}$/4$d$-4$\bar{d}$ variant of $\beta'$, and $\beta''$ phases in the 2H polytype.
                             The solid lines and circles represent the MLP and DFT predictions, respectively.
			\textbf{b} The MLP-predicted enthalpies of the four phases, relative to the $\alpha$ phase, as a function of pressure.
                           Note that the kink observed at 4 GPa in the $\beta'$ phase arises from a phase transformation between two distinct variants,
                           specifically from the 4$d$-4$\bar{d}$/4$d$-4$\bar{d}$ variant to the 4$d$-4$d$/4$\bar{d}$-4$\bar{d}$ variant.
			\textbf{c} Thermodynamic temperature-pressure phase diagram of In$_2$Se$_3$ predicted using MLP
                             through the thermodynamic integration method and the reversible scaling method.
                              The circles indicate the MLP-predicted phase boundary, while the triangle~\cite{van1975} and square~\cite{APL_pressure_Rasmussen2013}
                              represent the experimentally determined critical transition points.
	}
	\label{Figure2}
\end{figure}

The experimentally observed $\beta''$-In$_2$Se$_3$ also originates from the special arrangement of the central Se atoms,
accompanied by in-plane and out-of-plane displacements of outer-layer Se and In atoms (Fig.~\ref{Figure1}g).
This leads to the typical zig-zag striped morphology with a 2$\times\sqrt{3}$ surface unit cell observed
from scanning tunneling microscopy (STM)~\cite{ACSNano_Zhang2019,AdvancedScience_Chen2021,AM_electric_Zhang2022,NC_beta1_beta2_electric_field_Zhang2024}.
Following the experimental findings in Ref.~\cite{ACSNano_Zhang2019},
we successfully identified the $\beta''$-1T structure by quenching the $\beta'$-1T phase
to a low temperature of 77 K in our MD simulations, followed by structural relaxation and symmetry determination.
The resulting $\beta''$-1T structure exhibits nearly identical zig-zag striped morphology
with the structure proposed in Refs.~\cite{ACSNano_Zhang2019,AdvancedScience_Chen2021,NC_beta1_beta2_electric_field_Zhang2024}  (Supplementary Fig.~7g, i).
Moreover, both structures are dynamically stable (Supplementary Fig.~7k, m).
However, they exhibit distinct orientations of the central Se atoms due to the Mexican-hat-like PES (Supplementary Fig.~7a-f).
Importantly, the structure identified in this work demonstrates a lower energy by 11 meV per formula unit (meV/f.u.).
We also performed STM simulations and found that our identified structure
agree better with the experimental STM observation of Chen \emph{et al.}~\cite{AdvancedScience_Chen2021} (Supplementary Fig.~7j, n),
whereas the literature structure aligns better with the experimental STM observation of Zhang \emph{et al.}~\cite{NC_beta1_beta2_electric_field_Zhang2024} (Supplementary Fig.~7h, l).
These findings collectively suggest that both types of structures may coexist in experiments.
Besides the $\beta''$-1T polytype, the $\beta''$-2H and $\beta''$-3R phases
were also identified (Fig.~\ref{Figure1}g, Supplementary Fig.~5k),
and confirmed to be dynamically stable (Supplementary Fig.~4b, Supplementary Fig.~6k).

\begin{figure*}[ht!]
	\centering
\includegraphics[width=0.8\textwidth]{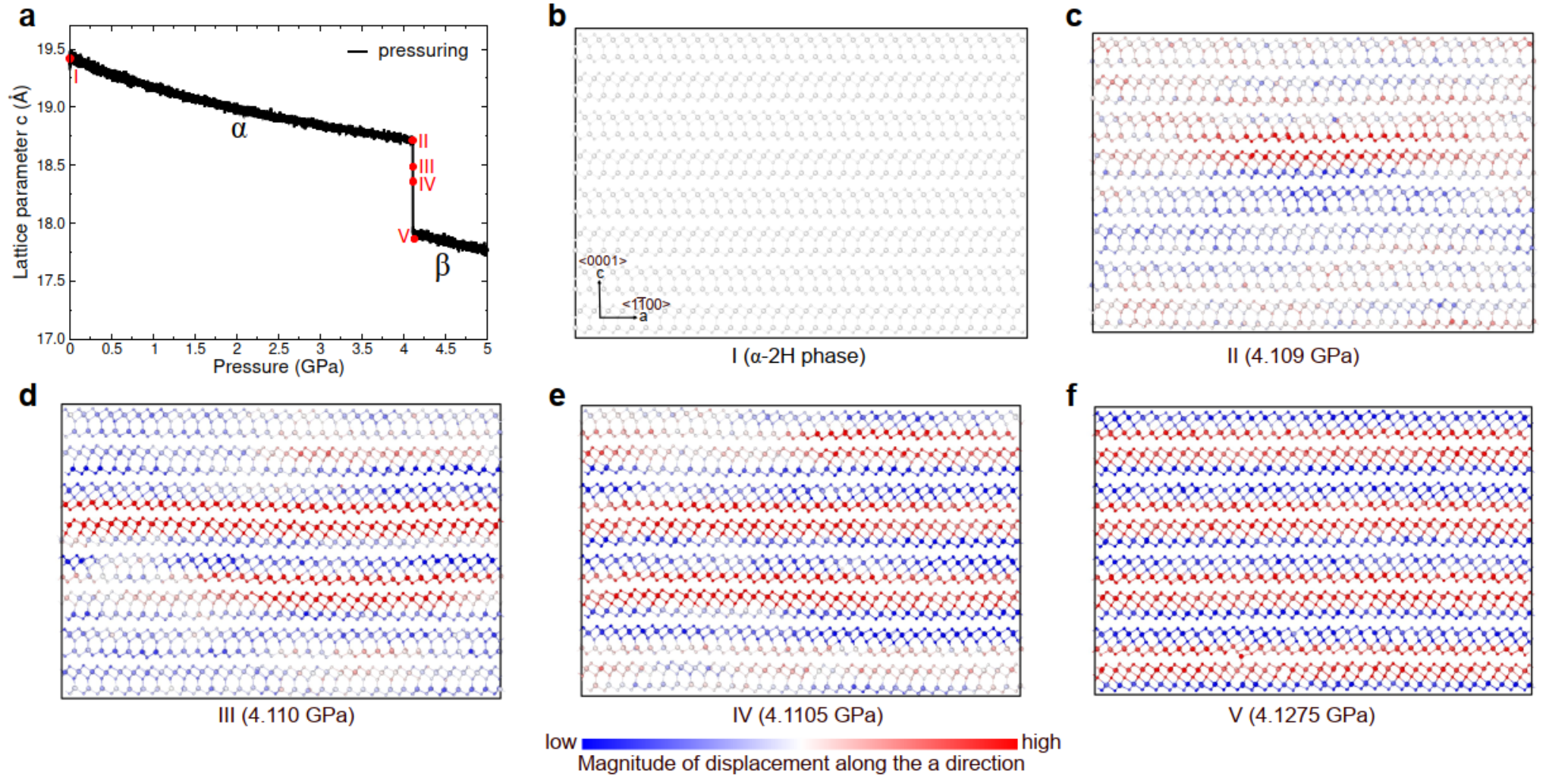}
		\caption{\textbf{| Pressure-induced phase transition from $\alpha$-2H to $\beta$-2H revealed by MD simulations.}
			\textbf{a} Evolution of the lattice parameter $c$ during pressurization (pressurization  rate: 0.67 GPa/ns) at 300 K.
            The initial $\alpha$-2H supercell including 2,560 atoms is constructed by replicating the $\alpha$-2H unit cell using the transformation matrix [16~16~0; $-$2~2~0; 0 0 4].
   			\textbf{b}-\textbf{f} MD snapshots corresponding to the simulation times marked by red dots in \textbf{a}.
            The large and small balls indicate the In and Se atoms, respectively.
            The color represents the magnitude of displacement along the $a$ direction relative to the $\alpha$-2H phase.
	}
	\label{Figure3}
\end{figure*}

Before proceeding to the next section, it is important to emphasize that
while the above discussions primarily centered on the 2H polytype,
the insights presented are broadly applicable to the 1T and 3R polytypes as well.
For instance, Supplementary Fig.~5 showcases several identified polymorphs with the 1T or 3R structures,
whose predicted lattice parameters and available experimental values are provided Supplementary Table 3.
The dynamical stability of these polymorphs has been confirmed in Supplementary Fig.~6.
Given the existence of multiple polytypes and the Mexican-hat-like PES in vdWs In$_2$Se$_3$,
it is plausible to expect the discovery of additional novel metastable phases beyond those identified in this study.

\begin{figure*}[ht!]
	\centering
\includegraphics[width=0.9\textwidth]{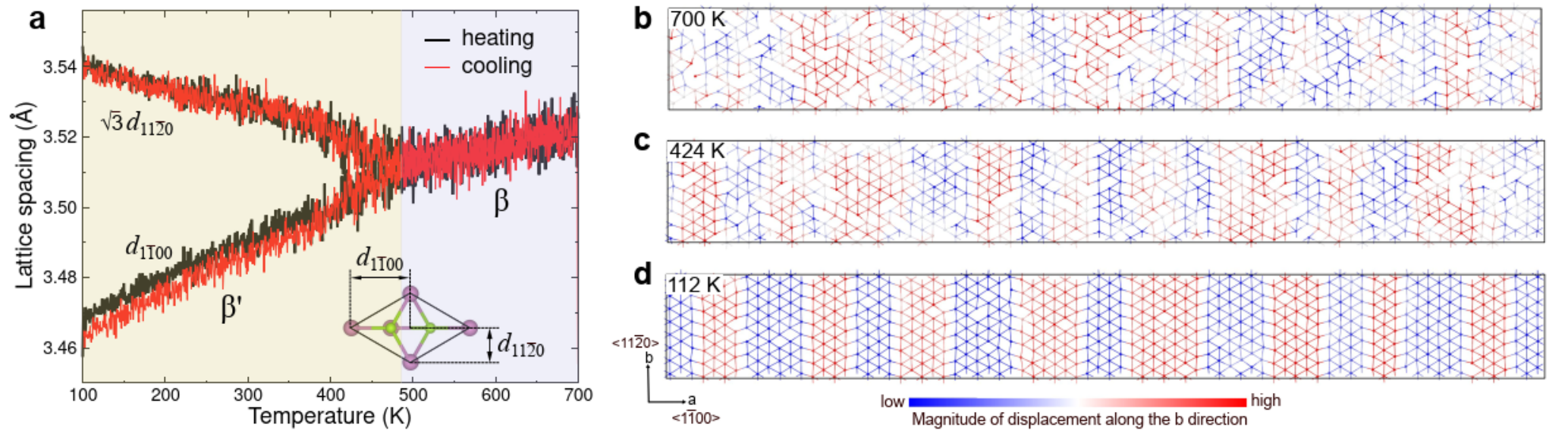}
		\caption{\textbf{| Temperature-induced reversible phase transitions between $\beta'$-2H and $\beta$-2H revealed by MD simulations.}
			\textbf{a} Evolution of the lattice spacings ($\sqrt{3}d_{11\bar{2}0}$ and $d$) during heating or cooling (heating/cooling rate: 0.04 K/ps) at 0 GPa.
                           The insert shows a top view of the local structure with the lattice spacings indicated. Large purple and small green spheres represent In and Se atoms, respectively.
            The initial $\beta$-2H supercell is constructed by replicating the $\beta$-2H unit cell using the transformation matrix [40~40~0; $-$8~8~0; 0 0 1].
            The initial $\beta'$-2H supercell is built by replicating the unit cell of the 4$d$-4$\bar{d}$/4$d$-4$\bar{d}$ variant in a $10\times8\times1$ configuration.
            Both supercells consist of 6,400 atoms.
			\textbf{b}-\textbf{d} MD snapshots during the cooling process at 700 K, 424 K, and 112 K, respectively.
            Only the central Se atoms in one quintuple layer are shown to better visualize the phase transitions.
            The color represents the magnitude of displacement along the $b$ direction relative to the ideal undistorted $\beta$-2H phase.
	}
	\label{Figure4}
\end{figure*}

\vspace{3mm}
\noindent\textbf{Thermodynamic phase diagram}\\

Having extensively explored the vast polymorphic phase space of vdWs In$_2$Se$_3$ and unveiled numerous novel phases,
we now turn to the study of their thermodynamics.
Here, we focus only on the experimentally observed polymorphs, specifically the $\alpha$, $\beta$, $\beta'$, and $\beta''$ polymorphs in the 2H polytype.
Fig.~\ref{Figure2}a presents the energy-volume curves for the four phases at 0K, as predicted by both MLP and DFT,
demonstrating excellent agreement between the two methods.
Consistent with experimental observations, the $\alpha$ phase displays the lowest equilibrium energy at 0 K, followed sequentially by the $\beta''$, $\beta'$, and $\beta$ phases.
Additionally, the $\alpha$ phase exhibits the largest equilibrium volume
and transforms into the $\beta'$ under increased pressure of up to 1.25 GPa at 0 K (Fig.~\ref{Figure2}b).
Further increasing the pressure to 5.10 GPa drives the transition from the $\beta'$ to the $\beta$ phase (Fig.~\ref{Figure2}b).

By incorporating the anharmonic effects through MD simulations,
we construct an \emph{ab initio} temperature-pressure phase diagram of In$_2$Se$_3$ using
the thermodynamic integration method~\cite{UnderstandingMolecularSimulation2002}
and the reversible scaling method~\cite{dekoningOptimizedFreeEnergyEvaluation1999} (see  "Methods").
As depicted in Fig.~\ref{Figure2}c, the $\alpha$ phase stabilizes in the region of low temperature and low pressure,
while the $\beta$ phase forms under conditions of high temperature and high pressure.
In the intermediate region between the $\alpha$ and $\beta$ phases, the $\beta'$ phase is found to be energetically more stable.
Specifically, at ambient pressure, the $\alpha$ phase undergoes transformation to the $\beta$ phase at an elevated temperature of 683 K
which aligns reasonably with the experimental result of 473 K~\cite{van1975}.
Additionally, the critical transition temperature decreases as the pressure increases.
At room temperature, the computed critical pressure for the $\alpha$ to $\beta$ transition is about 0.9 GPa,
showing good agreement with the experimental value of 0.7 GPa~\cite{APL_pressure_Rasmussen2013}.
Below room temperature, as the pressure increases, the $\alpha$ phase first transform into the intermediate $\beta'$ phase and then into the $\beta$ phase.
We note that the experimentally observed $\beta''$ polymorph is a metastable phase,
exhibiting a higher Gibbs free energy compared to the other three phases across the considered temperature and pressure range (Fig.~\ref{Figure2}b, Supplementary Fig.~8),
which explains its absence in the phase diagram.

\vspace{3mm}
\noindent\textbf{Dynamical polymorphic phase transitions}\\

After establishing the thermodynamic phase diagram,
we proceed to investigate the phase transitions among the $\alpha$, $\beta$, $\beta'$, and $\beta''$ polymorphs of the 2H polytype
under varying temperature, pressure, and strain conditions through MLP-accelerated MD simulations.
Our goal is to uncover the underlying microscopic mechanisms by elucidating the dynamical details of these phase transitions.

\emph{\textbf{$\alpha$ to $\beta$ phase transition.}}
The phase transformation from the $\alpha$ to $\beta$ phases is isosymmetric.
 It is commonly believed that the transformation involves a thermally activated interlayer shear glide,
during which the outer Se atoms migrate into interstitial sites, followed by an intralayer rearrangement~\cite{APL_pressure_Ke2014,AM_LaserLi2022}.
This process results in a compression of the interlayer distance.
Our room-temperature MD simulations successfully reproduce the pressure-induced first-order $\alpha$-$\beta$ phase transition,
as evidenced by the abrupt decrease in the lattice parameter $c$ at about 4 GPa in Fig.~\ref{Figure3}a.
The discrepancy between the critical transition pressure obtained from direct MD simulations and the thermodynamic value highlights the significant hysteresis.
In addition to pressure, the temperature-driven $\alpha$-$\beta$ phase transition is also well captured through heating MD simulations (Supplementary Fig.~9).
Analysis of the MD trajectory reveals that the phase transformation initiates with nucleation in the $ab$ plane (Fig.~\ref{Figure3}c),
and subsequently progresses in a layer-by-layer fashion along the $c$ axis (Fig.~\ref{Figure3}d-f).
Notably, this mechanism is similar to that observed in the phase transition of Ti$_3$O$_5$~\cite{Liu2024}.
Both our pressure- and temperature-dependent MD simulations confirm the
established "interlayer shear glide" mechanism~\cite{APL_pressure_Ke2014,AM_LaserLi2022} (Fig.~\ref{Figure3}b-f, Supplementary Fig.~9).
In accordance with our previous HAADF-STEM findings~\cite{Zhibo_SF_2024},
our MD simulations revealed the presence of stacking faults in the $\beta$-2H phase
when the simulation cell length was extended along the $c$ axis (Supplementary Fig.~10).

\begin{figure*}[ht!]
	\centering
\includegraphics[width=0.8\textwidth]{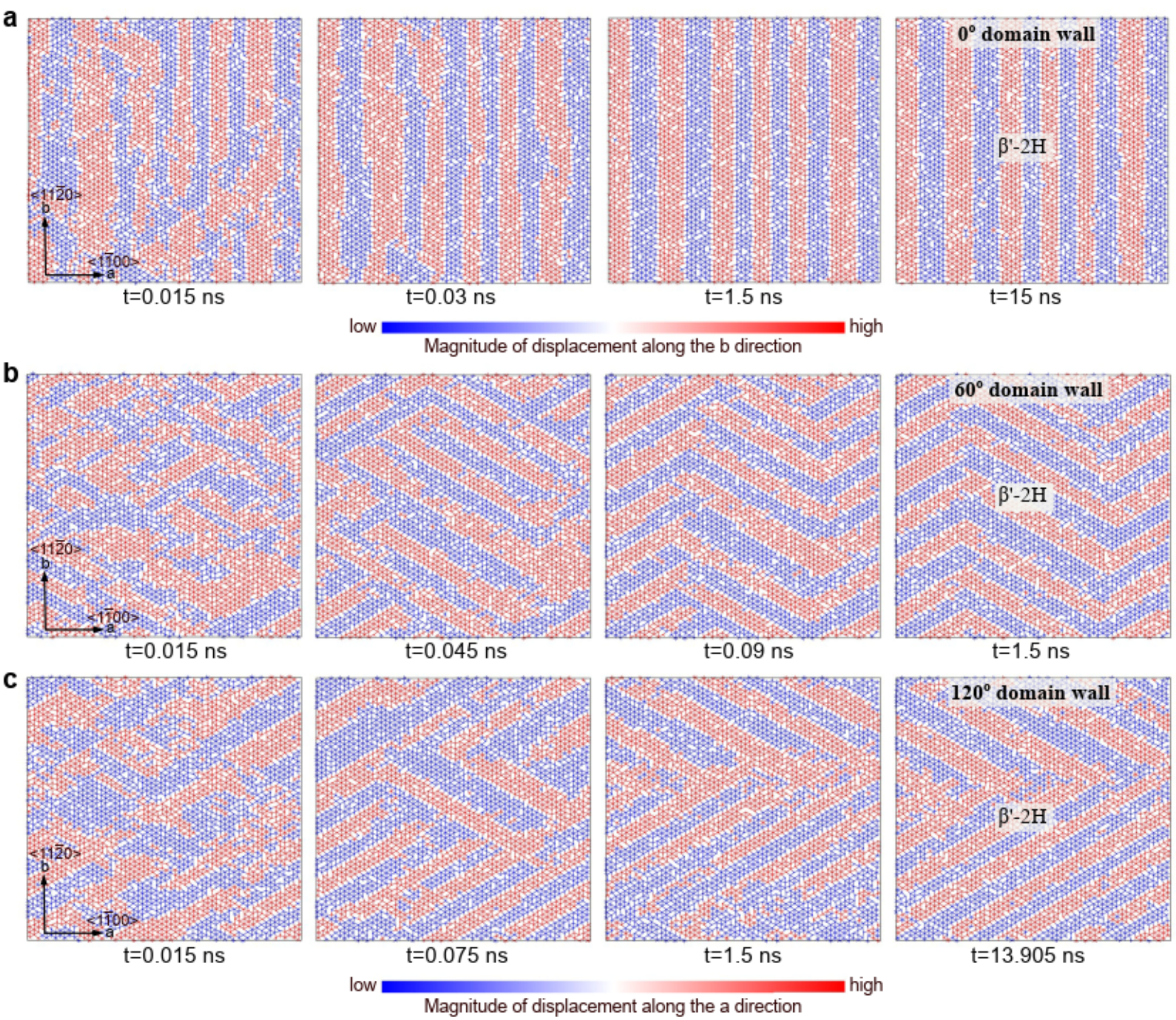}
		\caption{\textbf{| Domain walls in $\beta'$-2H revealed by MD simulations.}
			\textbf{a}-\textbf{c} Snapshots from three independent MD simulations at 300 K and 0 GPa,
             all starting from the same orthogonal $\beta$-2H supercell comprising 30,000 atoms but using different random seeds.
            The initial supercell  is constructed by replicating the $\beta$-2H unit cell using the  transformation matrix [30~30~0; $-$50~50~0; 0 0 1].
            Only the central Se atoms in one quintuple layer are shown to better visualize the domain walls.
            The colors in \textbf{a} and \textbf{b}-\textbf{c} represent the magnitude of displacement along the $b$ and $a$ directions, respectively,
            relative to the ideal undistorted $\beta$-2H phase.
	}
	\label{Figure5}
\end{figure*}

It is worth noting that a different mechanism for the electrical current-driven $\alpha$ to $\beta$ phase transition
has been identified by two recent studies~\cite{NC_JiWei_Wu2024,NM_Zhang2025}.
This mechanism is characterized by intralayer splitting and interlayer reconstruction,
which is observed exclusively in the 2H polytype but absent in the 3R polytype~\cite{NC_JiWei_Wu2024,NM_Zhang2025}.
Nevertheless, such a phenomenon is not observed in any of our MD simulations.
DFT calculations indicate that this mechanism is more energy-efficient than the "interlayer shear glide" mechanism~\cite{NC_JiWei_Wu2024,NM_Zhang2025}.
However, it is important to note that this conclusion is derived from the collective mechanism governing the
$\alpha$ to $\beta$ phase transition, which necessitates simultaneous gliding between two adjacent vdWs-gapped layers.
In contrast, we find that the phase transition from $\alpha$ to $\beta$
proceeds through a kinetically favorable nucleation mechanism, as displayed in Fig.~\ref{Figure3}c.
Since the intralayer split structures reported in Refs.~\cite{NC_JiWei_Wu2024,NM_Zhang2025} were included in our training dataset,
we conducted a computational experiment using MLP.
Specifically, we intentionally split a Se-In bond, causing the In atom to move into the vdWs gap---the step with the highest energy barrier---and then performed constant-temperature (700 K) and ambient-pressure MD simulations. Interestingly, the intralayer did not further split; instead, the In atom returned to the intralayer and re-formed the Se-In bond.
This computational experiment suggests that the intralayer splitting is unlikely under these conditions.
However, an important factor not yet addressed is the role of the electrical current applied in experiments~\cite{NC_JiWei_Wu2024,NM_Zhang2025}.
Whether it plays a dominant role in facilitating the intralayer splitting remains unclear, and we leave this question for future investigation.

\emph{\textbf{$\beta$ to $\beta'$ phase transition.}}
Experimentally, the $\beta$-$\beta'$ phase transitions are thermally driven and reversible~\cite{PRL_AFE2020,NC_ferroelasticity_Xu2021}.
Our large-scale heating and cooling MD simulations at ambient pressure accurately reproduce these reversible phase transitions,
which are manifested by the evolution of the lattice spacings $\sqrt{3}d_{11\bar{2}0}$ and $d_{1\bar{1}00}$ (see Fig.~\ref{Figure4}a).
Notably, the evolution of the lattice spacings during heating and cooling runs exhibits nearly overlapping trajectories, indicating small thermal hysteresis.
The excellent agreement between our MD simulations and experimental results~\cite{NC_ferroelasticity_Xu2021}
underscores the reliability of our MLP in capturing the phase transition dynamics.
By analyzing the MD trajectories, the dynamical details of the phase transition can be elucidated.
Taking the cooling process as an illustrative example, at a high temperature of 700 K,
thermally induced disorder is observed in the displacements of central Se atoms within the $\beta$ phase (Fig.~\ref{Figure4}b).
As the temperature drops below the critical point, the displacement of central Se atoms becomes increasingly ordered,
leading to the formation of the nanostripe-ordered $\beta'$ phase,
in which antiparallel displacements of central Se atoms emerge between adjacent nanostripes (Fig.~\ref{Figure4}c).
Further cooling to a low temperature of 100 K enhances the ordering of the $\beta'$ phase (Fig.~\ref{Figure4}d).
It is noteworthy, however, that the obtained $\beta'$ phase does not exhibit a perfect antiferroelectric structure with uniform nanostripe widths.
Instead, variations in nanostripe widths are observed, giving rise to net ferroelectricity.
This finding is consistent with prior experimental studies, which have also reported the presence of ferroelectricity in the
$\beta'$  phase~\cite{SA_Zheng2018,ACSNano_Zhang2019}.

\begin{figure*}[ht!]
	\centering
\includegraphics[width=0.85\textwidth]{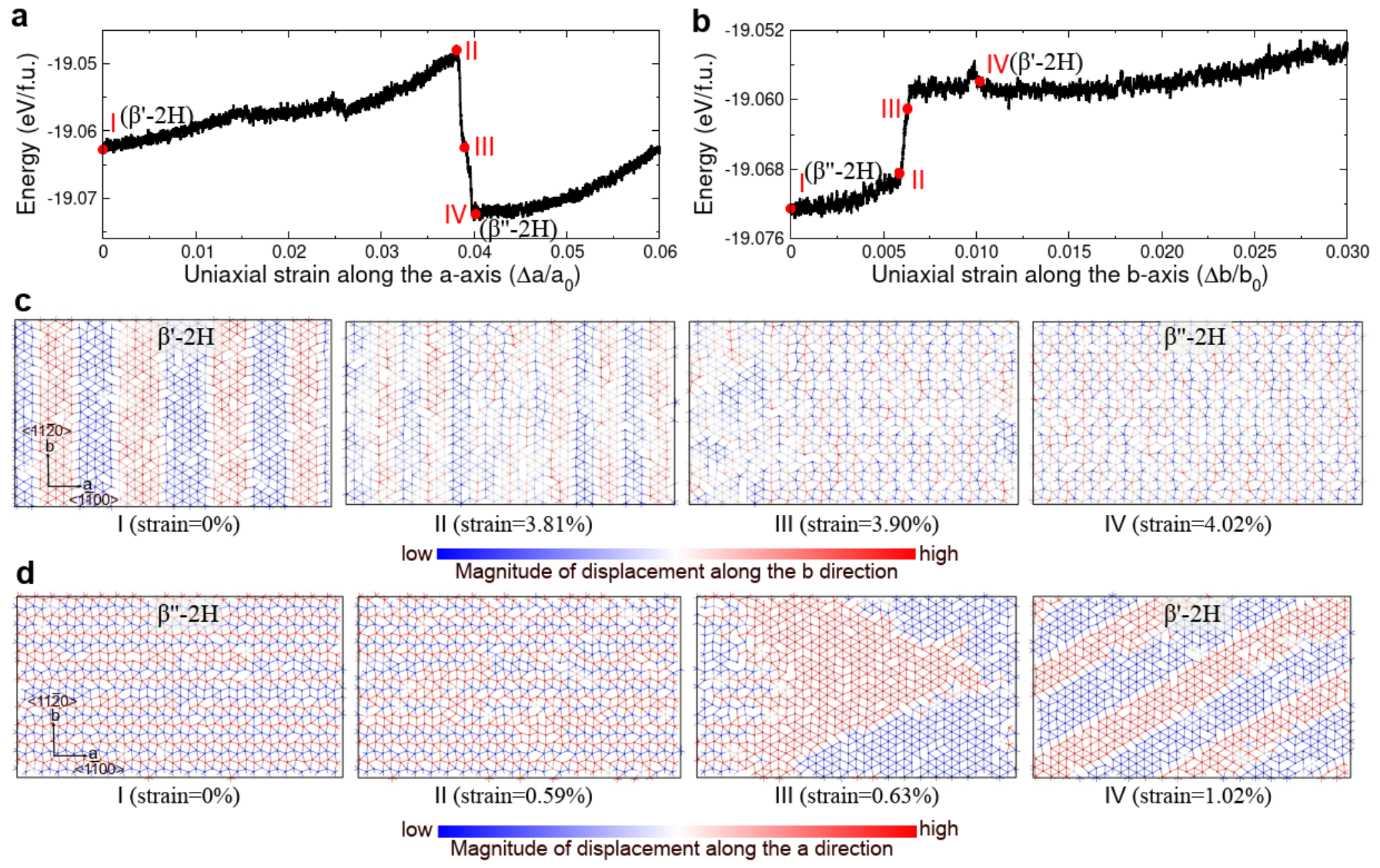}
		\caption{\textbf{| Strain-induced reversible phase transitions between $\beta'$-2H and $\beta''$-2H revealed by MD simulations.}
			\textbf{a} Evolution of the potential energy under uniaxial strain along the $a$-axis (strain rate: 0.01 ns$^{-1}$) at 300 K and 0 GPa, starting from the $\beta'$-2H phase.
			\textbf{b} Evolution of the potential energy under uniaxial strain along the $b$-axis (strain rate: 0.01 ns$^{-1}$) at 300 K and 0 GPa, starting from the $\beta''$-2H phase.
The initial $\beta'$-2H supercell is obtained by cooling the $\beta$-2H supercell including 8,000 atoms,
which is built by replicating the $\beta$-2H unit cell using the transformation matrix [20~20~0; $-$20~20~0; 0 0 1].
			\textbf{c} MD snapshots corresponding to the strain values marked by the red dots in \textbf{a}.
			\textbf{d} MD snapshots corresponding to the strain values marked by the red dots in \textbf{b}.
            Only the central Se atoms in one quintuple layer are shown to better visualize the phase transitions.
            The colors in \textbf{c} and \textbf{d} represent the magnitude of displacement along the $b$ and $a$ directions, respectively, relative to the ideal undistorted $\beta$-2H phase.
	}
	\label{Figure6}
\end{figure*}

We note that, in the $\beta'$ phase,
our MD simulations successfully captured not only the $0^\circ$ DW, but also the other two experimentally observed
DWs---namely, the $60^\circ$ and $120^\circ$ DWs~\cite{NC_ferroelasticity_Xu2021} (Fig.~\ref{Figure5}).
These three MD simulations were initiated from the same large supercell of the
$\beta$ phase, differing only in the initial random seeds employed within an isothermal-isobaric ensemble.
Despite these minor variations, all three DWs were ultimately obtained,
as they are symmetry-equivalent due to the hexagonal lattice of the parent $\beta$ structure~\cite{NC_ferroelasticity_Xu2021}.
It appears that the nucleation for a specific DW occurs very early in the simulation,
as shown in the first column of Fig.~\ref{Figure5}.
After this initial nucleation event, the growth of the DWs proceeds rapidly,
leading to a nearly complete phase transformation from the $\beta$ phase to the $\beta'$ phase within 1.5 ns.
This rapid transition highlights the dynamic nature of the phase change and
suggests that the nucleation process is a critical factor in the overall kinetics of the transformation.
Moreover, we observe the dynamical motion of the DWs,  resulting in variations in the width of nanostripes (Supplementary Fig.~11),

\emph{\textbf{$\beta'$ to $\beta''$ phase transition.}}
As previously discussed, the $\beta''$ phase is a metastable phase.
Experimentally, it is obtained by rapidly cooling the $\beta'$ phase to low temperatures~\cite{ACSNano_Zhang2019,AdvancedScience_Chen2021,AM_electric_Zhang2022,NC_beta1_beta2_electric_field_Zhang2024}.
This phase transition is reversible, and heating the $\beta'$ phase induces its transformation back into
the $\beta$ phase~\cite{ACSNano_Zhang2019,AdvancedScience_Chen2021,AM_electric_Zhang2022}.
Our MD simulations successfully reproduce these experimental observations (see Supplementary Fig.~12).
The transition temperature between the $\beta'$ and $\beta''$ phases, averaged over heating and cooling runs, is approximately 277 K,
in reasonable agreement with the experimentally determined value of 180 K~\cite{ACSNano_Zhang2019}.

In addition to temperature, we intriguingly find that uniaxial strain can also reversibly drive the $\beta'$ to $\beta''$ phase transitions,
as revealed by our unidirectional continuous tensile MD simulations (see Fig.~\ref{Figure6}).
Specifically, applying uniaxial strain along the $a$ axis to the $\beta'$ phase
induces nucleation of the $\beta''$ phase within the $\beta'$ matrix at 3.81\% strain (Fig.~\ref{Figure6}c, state II).
Increasing the strain further provides a sufficient driving force for the growth of the $\beta''$ phase (Fig.~\ref{Figure6}c, state III).
At a strain of 4.02\%, the $\beta'$ to $\beta''$ phase transition completes (Fig.~\ref{Figure6}c, state IV).
Conversely, for the reverse $\beta''$ to $\beta''$ phase transition,
the nucleation of the $\beta'$ phase begins when uniaxial strain applied to the $\beta''$ phase along the $b$ axis reaches 0.59\%  (Fig.~\ref{Figure6}d, state II).
With a slight increase in strain, the $\beta'$ phase rapidly grows (Fig.~\ref{Figure6}d, state III),
eventually forming a fully nanostripe-ordered $\beta'$ phase (Fig.~\ref{Figure6}d, state IV).
Note that the directional anisotropy in strain response originates from distinct in-plane lattice parameters of the $\beta'$ and $\beta''$ phases (Supplementary Table~1).
Importantly, the critical strains necessary for the phase transitions can be feasibly achieved under experimental conditions.

This strain-mediated reversible phase transition is particularly intriguing, as it accompanies the polarization modulations (Supplementary Fig.~13).
For example, applying a 1.5\% uniaxial strain along the $a$-axis to the $\beta'$-2H phase triggers an antiferroelectric-to-ferroelectric transition.
The ferroelectric polarization grows with increasing strain up to 4\%, beyond which the $\beta''$-2H phase forms and the polarization vanishes (Supplementary Fig.~13a).
Conversely, applying a 0.59\% uniaxial strain along the $b$-axis to the non-polar $\beta''$-2H phase induces ferroelectricity,
and further increasing the strain drives  the transformation into the antiferroelectric $\beta'$-2H phase (Supplementary Fig.~13b)
 By decoupling phase transitions from electric fields and enabling bidirectional control via strain,
our findings open avenues for designing novel strain-programmable electronic devices,
such as nonvolatile memories or tunable sensors~\cite{AM_Review_Li2024,Nanophotonics2024}.
We note that, akin to previously studied phase transitions, the in-plane nucleation and
$c$-axis layer-by-layer growth mechanism is also relevant to the strain-induced $\beta'$ to $\beta''$ phase transition (see Supplementary Fig.~14).
This observation underscores the broader applicability of such a mechanism in the phase transitions of vdWs layered materials.

\section*{Discussion}

The polymorphic phase transitions in vdWs In$_2$Se$_3$ have long captivated the scientific community,
owing to their intricate atomic rearrangements and promising technological applications.
However, unraveling the atomic-scale mechanisms driving these transitions has been challenging,
hindered by the limitations of experimental techniques in achieving sufficient spatial and temporal resolution,
as well as the restricted time and length scales of \emph{ab initio} simulations.
Our work addresses these challenges by MLP-accelerated MD simulations,
unlocking crucial insights into the atomic-scale dynamics of phase transitions in vdWs In$_2$Se$_3$.

The developed moment tensor MLP demonstrates remarkable versatility,
reproducing key physical properties such as lattice parameters, energy-volume relationships, and phonon dispersions across 1T, 2H, and 3R polytypes,
while predicting thermodynamic temperature-pressure phase diagram with quantitative accuracy.
Importantly, the high efficiency of MLP enables large-scale MD simulations,
capturing the formation of nucleation, domain walls, and stacking faults---processes central to understanding phase transition pathways.
Our simulations reveal that the polymorphic transitions in vdWs In$_2$Se$_3$ proceed via kinetically favored in-plane nucleation, followed by layer-by-layer growth.
Moreover, a novel strain-induced reversible phase transformations between the $\beta'$ and $\beta''$ polymorphs
has been identified. The polarization modulations associated with reversible $\beta'$-$\beta''$ transitions
open promising pathways for the development of low-energy logic devices.

Beyond reproducing known polymorphic transitions, the MLP also enables the exploration of the vast polymorphic phase space of vdWs In$_2$Se$_3$,
uncovering numerous dynamically stable yet energetically competing phases.
Notably, we identify a new $\beta''$ phase,
which exhibits dynamical stability, aligns with experimentally observed zig-zag striped morphology,
and is energetically more favorable by 11 meV/f.u. as compared to previously reported structural model.
The structural flexibility and metastability of these phases arise from the layered nature of vdWs interactions,
flat imaginary phonon modes, and the Mexican-hat-like PES in the $\beta$ phase~\cite{APR_cal_Huang2021,NC_cal_Huang2024}.
The layered nature of vdWs In$_2$Se$_3$ inherently promotes stacking disorder and defect-mediated phase stability.
The unique PES topology, characterized by shallow energy barriers, facilitates spontaneous structural rearrangements
under thermal or mechanical perturbations, explaining the interconversion pathways among different polymorphs.
These findings suggest new avenues in designing novel vdW phase-change materials,
where metastable states and defect engineering could be leveraged for tailored functionality.
Future studies will focus on exploring the properties of these metastable phases beyond their structural and thermodynamic characteristics,
such as their electronic structure~\cite{NC_cal_Huang2024,Felton2025}, thermal conductivity~\cite{Zhou2016,ShiLiu2025},
and optical properties~\cite{AM_LaserLi2022,PhysRevB.98.165134}, to unlock their potential for practical applications.

\section*{Methods}

\noindent\textbf{First-principles calculations}\\
First-principles calculations were conducted using the Vienna \emph{ab initio} simulation package (VASP)~\cite{Kresse1996PRB}.
The generalized gradient approximation parameterized by Perdew-Burke-Ernzerhof (PBE)~\cite{Perdew1996PRL} was used for the exchange-correlation functional.
The projector augmented wave pseudopotentials~\cite{PhysRevB.50.17953,PhysRevB.59.1758} ({\tt In\_d} and {\tt Se}) were adopted.
The vdWs interactions were accounted for via the D3 method of Grimme \emph{et al.}~\cite{10.1063/1.3382344} with standard zero damping.
A plane wave cutoff of 450 eV and a $\Gamma$-centered $k$-point grid with a spacing of 0.19 \AA$^{-1}$ between $k$ points were employed,
which ensures the convergence of total energy better than 1 meV/atom.
The Gaussian smearing method with a smearing width of 0.05 eV was used.
The electronic optimization was performed until the total energy difference between two consecutive iterations was less than 10$^{-6}$ eV.
The structures were optimized until the forces were smaller than 1 meV/\AA.
The harmonic phonon dispersions were calculated using the Phonopy code~\cite{Togo2015-pho}.
The variable-cell climbing image nudged elastic band method~\cite{10.1063/1.3684549} was employed to estimate the energy barrier for the phase transition.
The STM simulation was conducted using the p4vasp code~\cite{p4vasp}.

\vspace{3mm}
\noindent\textbf{MLP construction}\\
For the on-the-fly active learning procedures based on the Bayesian linear regression~\cite{JinnouchiPRL2019,JinnouchiPRB2019},
the separable descriptors~\cite{doi:10.1063/5.0009491} based on the smooth overlap of atomic position~\cite{SOAP2013} were employed.
The cutoff radius for the three-body descriptors and the width of the Gaussian functions used
for broadening the atomic distributions of the three-body descriptors were set to 5~\AA~and 0.5~\AA, respectively.
The number of radial basis functions used to expand the radial descriptor was set to 10.
The parameters for the two-body descriptors were the same as those for the three-body descriptors.
To enhance the efficiency of MLP, the training dataset was refitted using the moment tensor potential (MTP).
For fitting the MTP, a cutoff radius of 6.0~\AA~was used, and the number of radial basis functions was set to be 8.
The basis sets were optimized by refining the contraction process of moment tensors,
which greatly speeds up the calculations but without losing the accuracy~\cite{wang2024}.
The basis functions of the MTP were selected such that the level of scalar basis did not exceed 26.
The maximum moment tensor order was set to 4.
The MTP incorporated up to five-body interactions, resulting in a total of 4,031 scalar basis functions.

\vspace{3mm}
\noindent\textbf{Molecular dynamics simulations}\\
Molecular dynamics simulations were performed using the LAMMPS code~\cite{Thompson2022}
interfaced with the moment tensor MLP~\cite{Novikov_2021}.
The simulations were carried out in an isothermal-isobaric (NPT) ensemble.
The temperature was controlled using a Nos\'e-Hoover thermostat~\cite{Hoover1985,Nose1991} with a relaxation time of 0.2 ps.
The pressure was controlled with the Parrinello-Rahman barostat~\cite{Parrinello1981} with a relaxation time of 0.1 ps.
The time step was set to 1.5 fs. The visualizations of MD snapshots and trajectories were generated using the OVITO software~\cite{ovito}.

\vspace{3mm}
\noindent\textbf{Thermodynamic phase diagram calculations}\\	
The thermodynamic phase diagram is calculated using the thermodynamic integration method~\cite{UnderstandingMolecularSimulation2002}
and the reversible scaling method~\cite{dekoningOptimizedFreeEnergyEvaluation1999}.
The MD simulations employed 400-atom supercells of the $\alpha$ phase and
the 4$d$-4$\bar{d}$/4$d$-4$\bar{d}$ variant of the $\beta'$ phase.
The equilibrium volume $V_0$ at the initial temperature $T_0$ and pressure $P$ was determined through a 5 ns NPT MD simulation.
For the $\alpha$ phase, the harmonic model was derived from analytical computation of the MTP potential energy hessian,
whereas the harmonic model for the $\beta'$ phase was obtained from the
stochastic self-consistent harmonic approximation (SSCHA)~\cite{monacelliStochasticSelfconsistentHarmonic2021} free energy Hessian at 100 K.
The harmonic free energy $A_\mathrm{har}$ is calculated as
\begin{equation}
	A_{\mathrm{har}}(V,T_{0})=k_{B}T_{0}\sum_{i=1}^{3N-3}\ln\frac{\hbar\omega_{i}}{k_{B}T_{0}},
\end{equation}
where $N$ denotes the total number of atoms in the system,
$\omega_i$ represents the $i$-th non-zero phonon frequency obtained from the harmonic model,
$k_{B}$ is the Boltzmann constant, and $\hbar$ is the reduced Planck constant.

The thermodynamic integration between the harmonic Hamiltonian ($\mathcal{H}_\text{har}$)
and the MTP Hamiltonian ($\mathcal{H}_\text{MTP}$) is performed using the following Hamiltonian:
\begin{equation}
	\mathcal{H}(\lambda) = \lambda(t) \mathcal{H}_\text{har} + (1 - \lambda(t))\mathcal{H}_\text{MTP},
\end{equation}
where the switching function $\lambda(t)$=$t^5(70t^4-315t^3+540t^2-420t+126)$~\cite{dekoningEinsteinCrystalReference1996}
controls the transition: $t$ ramps linearly from 0 to 1 (forward direction) and from 1 to 0 (reverse direction).
The Helmholtz free energy is computed by
\begin{equation}
	A_{cm}(V,T_{0}) = A_{\mathrm{har}}(V,T_{0}) + \frac{1}{2}\left( \int_{0}^{1}\frac{d\lambda}{dt} \Delta U \, dt - \int_{1}^{0}\frac{d\lambda}{dt} \Delta U \, dt \right),
\end{equation}
where $\Delta U = U_\text{MTP} - U_\text{har}$ represents the difference between the MTP potential and the harmonic potential.
The integration employed $10^6$ steps for both forward and reverse directions.
To account for the constraint on the system's center of mass,
a correction term was also applied~\cite{polsonFinitesizeCorrectionsFree2000}:
\begin{equation}
	 A(V, T_0) - A_{cm}(V, T_0) = -\frac{3}{2} k_B T \ln \left( \frac{V}{N} \frac{2\pi \hbar^2}{k_BTM} \right),
\end{equation}
where $M$ is the total mass of the system.

The Gibbs free energy at $T_0$ is then derived as~\cite{chengComputingAbsoluteGibbs2018}
\begin{equation}
	G(P, T_0) = A(V, T_0) + PV_0 + k_B T \ln \rho (\mathbf{h}|P,T_0),
\end{equation}
where the matrix $\mathbf{h}$ represents the simulation cell dimensions,
and $\rho (\mathbf{h}|P,T_0)$ quantifies the likelihood of sampling specific cell configurations $\mathbf{h}$ in the NPT ensemble at $P$ and $T_0$.
The temperature-dependent Gibbs free energy is obtained through the reversible scaling method~\cite{dekoningOptimizedFreeEnergyEvaluation1999}
\begin{equation}
	G(T) - G(T_0) = \int_{1}^{\tau_f} d\tau \langle U + PV \rangle,
\end{equation}
where $\tau$=$T/T_0$ is the reduced temperature, and $\tau_f$=$T_f / T_0$ with the final temperature $T_f$=700 K.

Note that for the $\beta$-$\beta'$ phase transition,
determining the transition temperature using the thermodynamic integration method is challenging due to the interchangeable nature of these two phases.
Therefore,  direct MD simulations were employed.
Specifically, for each pressure, five independent heating and five independent cooling MD simulations were conducted over a temperature range of 100 K to 700 K in 20 K increments.
At each temperature step, the system was equilibrated for 100 ps, followed by a 1 ns production run for structural sampling.
Structures were collected at a sampling rate of 0.1 ps.
Afterwards, the sampled structures were averaged, and the actual phase was determined.
The transition temperature was then determined as the mean of the transition temperatures obtained from all the heating and cooling MD simulations.

\section*{Data availability}
The data supporting the findings of this study, such as the predicted crystal structures, training and validation datasets,
and the developed machine learning potential, will be made available upon publication.

\section*{Code availability}
VASP is acquired from the VASP Software GmbH (see \href{https://www.vasp.at/}{www.vasp.at});
MLlP is available at \href{https://mlip.skoltech.ru/}{mlip.skoltech.ru};
LAMMPS is available at \href{https://www.lammps.org/}{www.lammps.org};
OVITO is available at \href{https://www.ovito.org/}{www.ovito.org};
Phonopy is available at \href{https://phonopy.github.io/phonopy/}{phonopy.github.io/phonopy};
p4vasp is available at \href{https://github.com/SSCHAcode}{github.com/SSCHAcode};
SSCHA is available at \href{https://github.com/orest-d/p4vasp}{github.com/orest-d/p4vasp}.

\bibliography{Reference}
\bibliographystyle{naturemag}

\section*{Acknowledgements}
This work is supported by the National Natural Science Foundation of China (Grants No.~52422112, No.~52401027, No.~52188101, and No.~52201030),
the Science and Technology Major Project of  Liaoning province (2024JH1/11700032),
the Liaoning Province Science and Technology Planning Project (2023021207-JH26/103 and RC230958),
the National Key R{\&}D Program of China 2021YFB3501503,
and the Special Projects of the Central Government in Guidance of Local Science and Technology Development (2024010859-JH6/1006).

\section*{Author contributions}
P.L. conceived the project.
P.L. designed the research with the help of X.-Q.C..
P.L. and X.-Q.C. supervised the project.
M.L., J.W., and P. L. performed the calculations.
Q.W., Z.L., and Y.S. contributed to discussions.
P.L. wrote the manuscript with inputs from other authors.
All authors comments on the manuscript.
M.L., J.W., and P.L. contributed equally to this work.

\section*{Competing interests}
The authors declare no competing interests.

\section*{Additional information}
Supplementary information is available.
Correspondence and requests for additional materials should be addressed to P.L. or X.-Q.C..

\end{document}


\captionsetup[figure]{labelfont={bf},labelformat={default},labelsep=period,name={Supplementary Fig.}}
\captionsetup[table]{labelfont={bf},labelformat={default},labelsep=period,name={Supplementary Table}}

		\maketitle
		\vspace{-5mm}
		\begin{center}
			\begin{minipage}{1\textwidth}
				\begin{center}
					\textit{
						\textsuperscript{1} Shenyang National Laboratory for Materials Science, Institute of Metal Research, Chinese Academy of Sciences, 110016 Shenyang, China
						\\\textsuperscript{2} School of Materials Science and Engineering, University of Science and Technology of China, Shenyang 110016, China
						\vspace{5mm}
						\\{$\star$} These authors contribute equally.
						\\{$\dagger$} Corresponding to: ptliu@imr.ac.cn, xingqiu.chen@imr.ac.cn
						\vspace{5mm}
					}
				\end{center}
			\end{minipage}
		\end{center}
		
		\setlength\parindent{13pt}

	\newpage
	\clearpage

\section*{Supplementary Figures}
\vspace{6mm}

\begin{figure*}[!h]
		\centering  \includegraphics{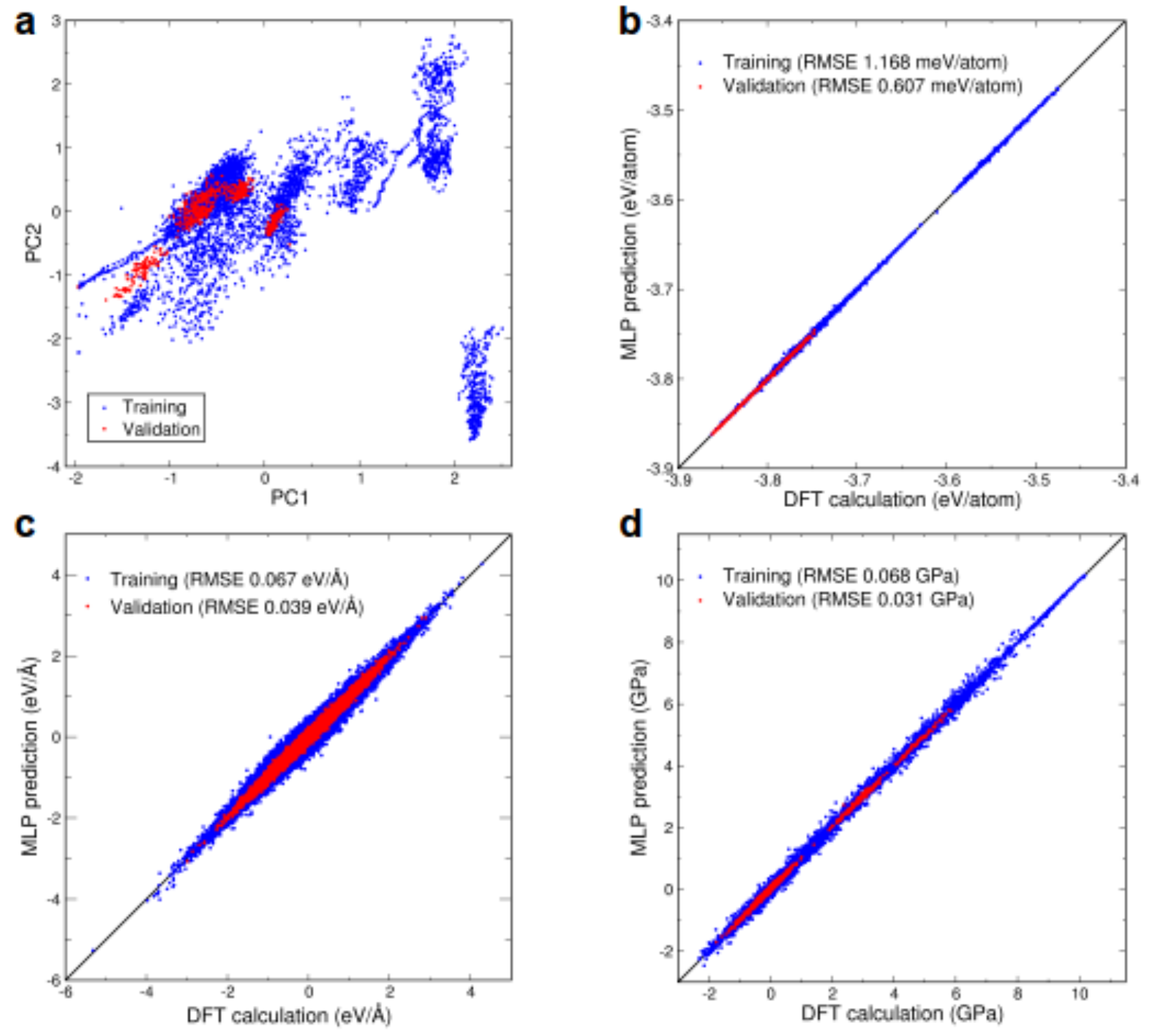}
		\caption{\textbf{| Training and validation datasets and machine learning potential (MLP) predictions vs. density functional theory (DFT) results.}
           \textbf{a}, The kernel principal component analysis map of training and validation structures.
			\textbf{b}, MLP predicted energies vs. DFT results.
			\textbf{c}, MLP predicted forces vs. DFT results.
			\textbf{d}, MLP predicted stress tensors vs. DFT results.
The training and validation data points are represented by blue and red dots, respectively.
The root-mean-square errors (RMSEs) for energies, forces, and stress tensors are provided for both the training and validation datasets.
Note that the validation dataset covers only a fraction of the phase space represented by the training dataset,
thereby resulting in significantly smaller RMSEs for energies, forces, and stress tensors in the validation dataset compared to the training dataset.
             }
		\label{fig:FigS1}
	\end{figure*}

\newpage
\clearpage
\begin{figure*}[ht!]
	\centering
\includegraphics[width=0.95\textwidth]{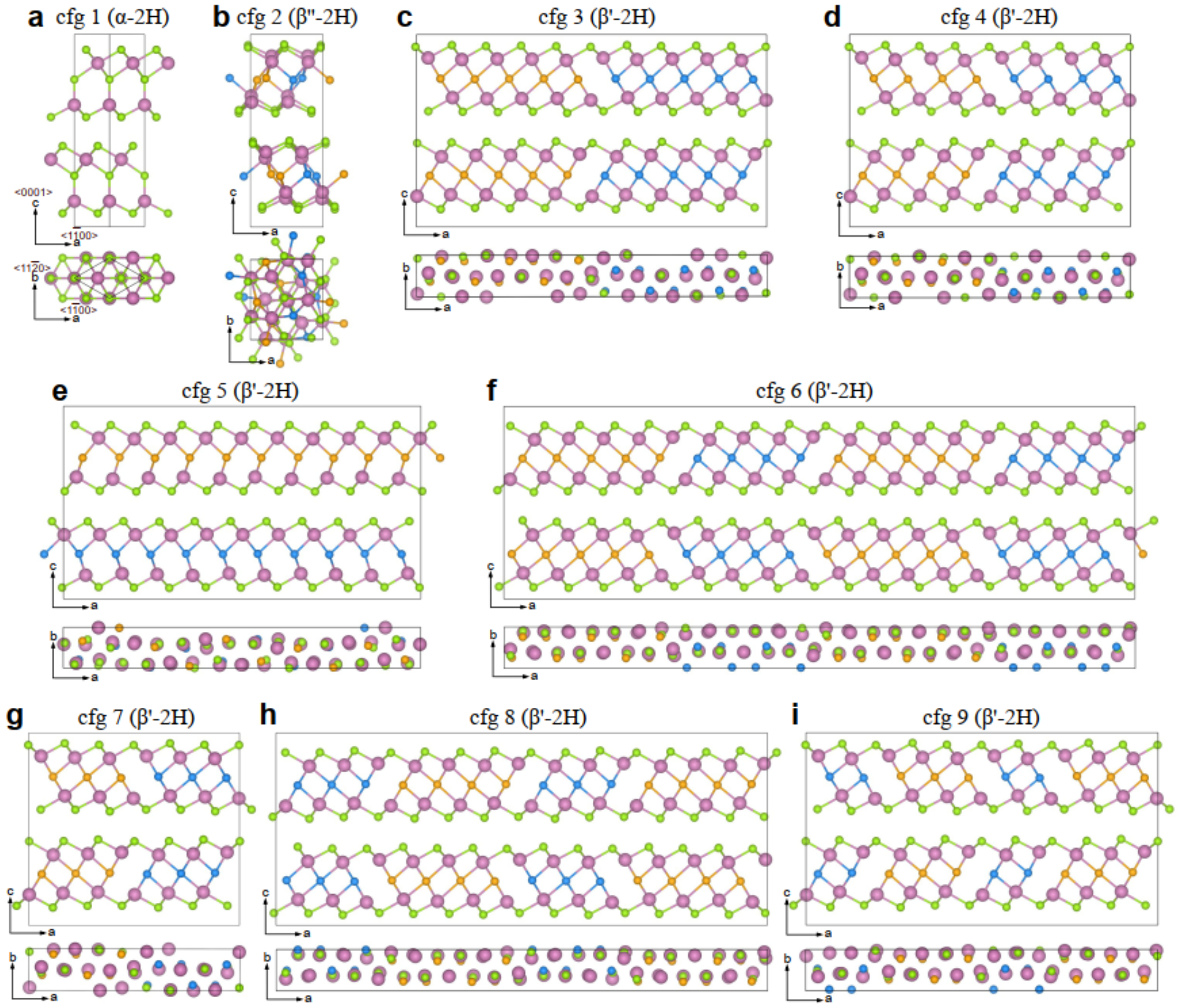}
		\caption{\textbf{| Identified polymorphs of van der Waals (vdWs) In$_2$Se$_3$ in the 2H polytype.}
\textbf{a} cfg1: $\alpha$-2H.
\textbf{b} cfg2: $\beta''$-2H.
\textbf{c} cfg3: The 5$d$-5$\bar{d}$/5$d$-5$\bar{d}$  variant of $\beta'$-2H.
\textbf{d} cfg4: The 4$d$-4$\bar{d}$/4$d$-4$\bar{d}$  variant of $\beta'$-2H.
\textbf{e} cfg5:  The 10$d$/10$\bar{d}$  variant of $\beta'$-2H.
\textbf{f} cfg6: The 5$d$-4$\bar{d}$/5$d$-4$\bar{d}$ variant of $\beta'$-2H.
\textbf{g} cfg7: The 3$d$-3$\bar{d}$/3$d$-3$\bar{d}$ variant of $\beta'$-2H.
\textbf{h} cfg8: The 4$d$-3$\bar{d}$/4$d$-3$\bar{d}$ variant of $\beta'$-2H.
\textbf{i} cfg9: The 3$d$-2$\bar{d}$/3$d$-2$\bar{d}$ variant of $\beta'$-2H.
	}
	\label{fig:FigS2}
\end{figure*}

\begin{figure*}[ht!]
	\centering
\ContinuedFloat 
\includegraphics[width=0.95\textwidth]{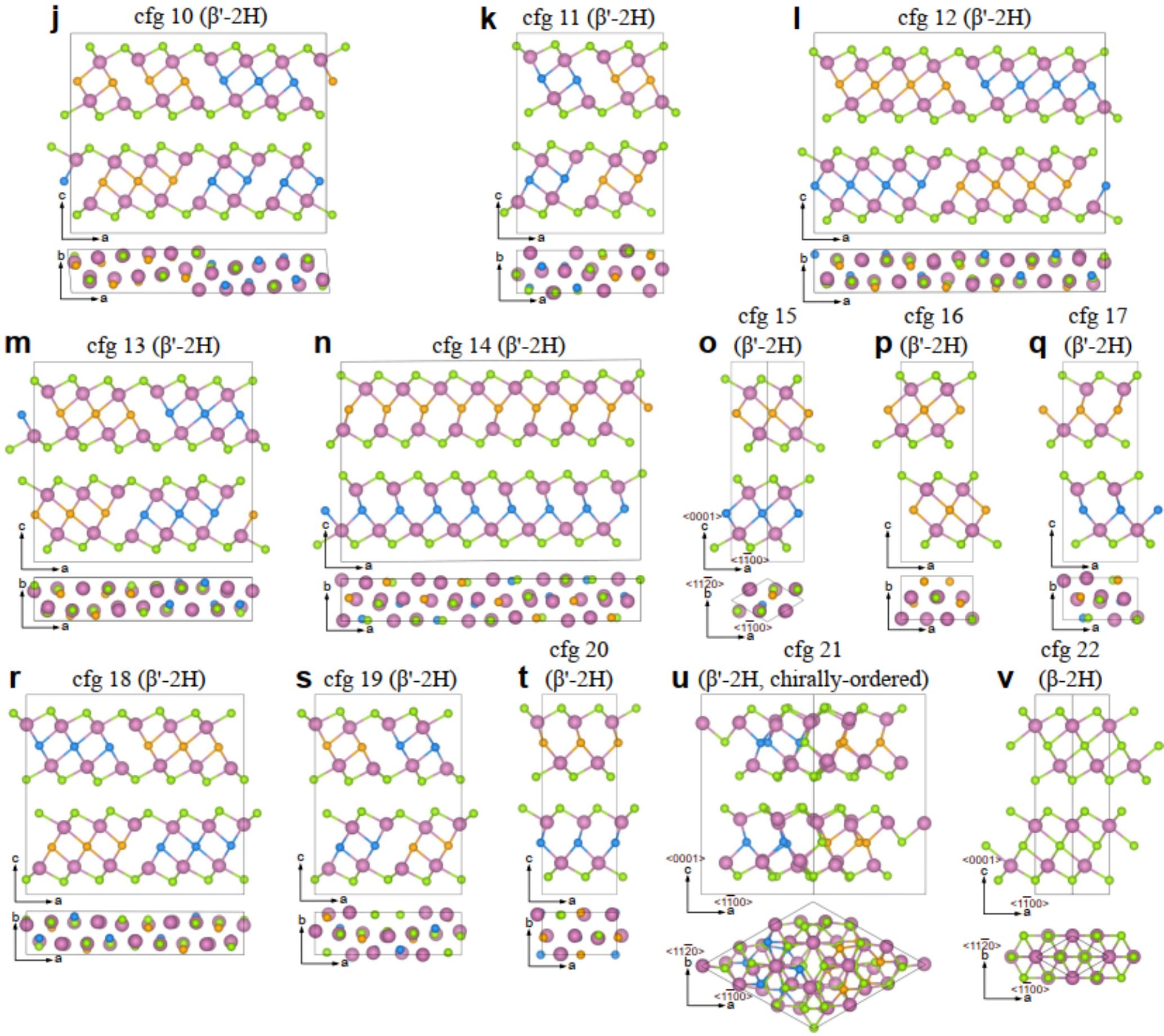}
		\caption{\textbf{| Continued.}
\textbf{j} cfg10: The 4$d$-3$\bar{d}$/3$d$-4$\bar{d}$ variant of $\beta'$-2H.
\textbf{k} cfg11: The 2$d$-2$\bar{d}$/2$d$-2$\bar{d}$ variant of $\beta'$-2H.
\textbf{l} cfg12: The 4$d$-4$\bar{d}$/4$\bar{d}$-4$d$ variant of $\beta'$-2H.
\textbf{m} cfg13: The sheared 3$d$-3$\bar{d}$/3$d$-3$\bar{d}$ variant of $\beta'$-2H. Note that cfg13 differs with cfg7 in the stacking.
\textbf{n} cfg14:  The 8$d$/8$\bar{d}$ variant of $\beta'$-2H.
\textbf{o} cfg15: The hexagonal 2$d$/2$\bar{d}$ variant of $\beta'$-2H.
\textbf{p} cfg16: The 2$d$/2$d$ variant of $\beta'$-2H.
\textbf{q} cfg17: The orthogonal 2$d$/2$\bar{d}$ variant of $\beta'$-2H [Space group: $Pna2_1$ (33)].
\textbf{r} cfg18: The 3$d$-3$\bar{d}$/3$\bar{d}$-3$d$ variant of $\beta'$-2H.
\textbf{s} cfg19: The 2$d$-2$\bar{d}$/2$\bar{d}$-2$d$ variant of $\beta'$-2H.
\textbf{t} cfg20: The orthogonal 2$d$/2$\bar{d}$ variant of $\beta'$-2H [Space group: $Pca2_1$ (29)].
                          Note that cfg20 differs with cfg17 in the polar displacement of central Se atoms, yielding different space groups.
\textbf{u} cfg21: The polymorph with chirally-ordered central Se atoms.
\textbf{v} cfg22: $\beta$-2H.
Note that the large balls indicate the In atoms, whereas the small balls denote the Se atoms.
The orange and blue small balls represent the antiparallel displacements of central Se atoms along the $\langle11\bar{2}0\rangle$ direction,
except for the case in \textbf{u} where they are just used to guide the eye for the chiral displacements of specified central Se atoms.
The notation $n_1d$-$n_2\bar{d}$/$n_3d$-$n_4\bar{d}$ indicates that
the the upper and bottom quintuple layers consist of $n_1d$-$n_2\bar{d}$ and $n_3d$-$n_4\bar{d}$ structural motifs, respectively.
Similarly, the notation $md$/$n\bar{d}$ represents that the the upper and bottom quintuple layers consist of $md$ and $n\bar{d}$ structural motifs, respectively.
Here, $\bar{d}$ and $d$ denote that the central Se atoms with a width of $d_{1\bar{1}00}$
displace along the $\langle11\bar{2}0\rangle$ and $\langle\bar{1}\bar{1}20\rangle$ directions, respectively.
	}
	\label{fig:FigS2}
\end{figure*}

\newpage
\clearpage
\begin{figure*}[ht!]
	\centering
\includegraphics[width=0.7\textwidth]{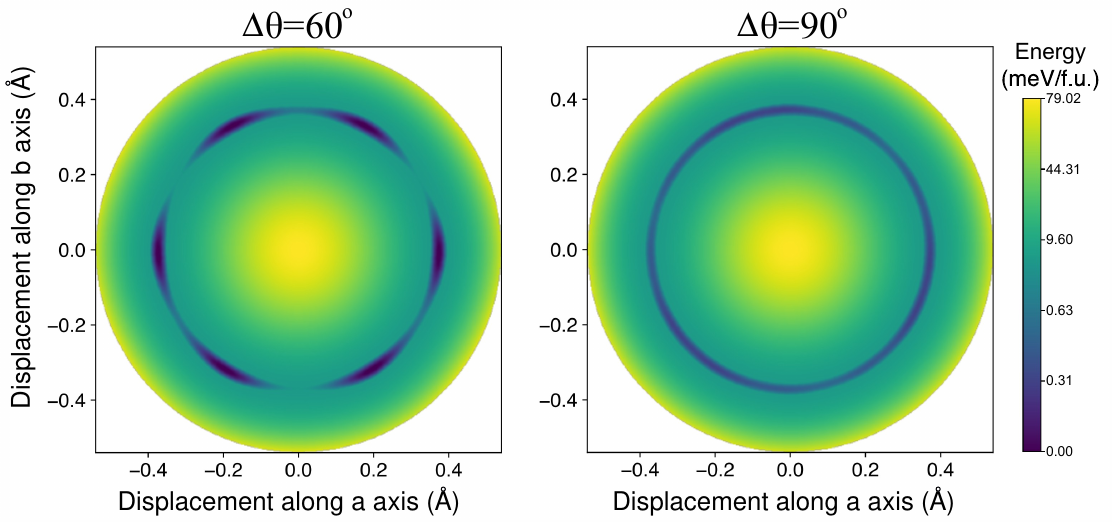}
		\caption{\textbf{| The Mexican-hat-like potential energy surface (PES) for the middle-layer Se atoms in $\beta$-2H.}
The Mexican-hat-like PES when adjusting the positions of the central Se atoms in both quintuple layers of $\beta$-2H
for two phase difference angles: $\phi$=$60\degree$ and $\phi$=$90\degree$.
	}
	\label{fig:FigS3}
\end{figure*}

\newpage
\clearpage
\begin{figure*}[ht!]
	\centering
\includegraphics[width=0.95\textwidth]{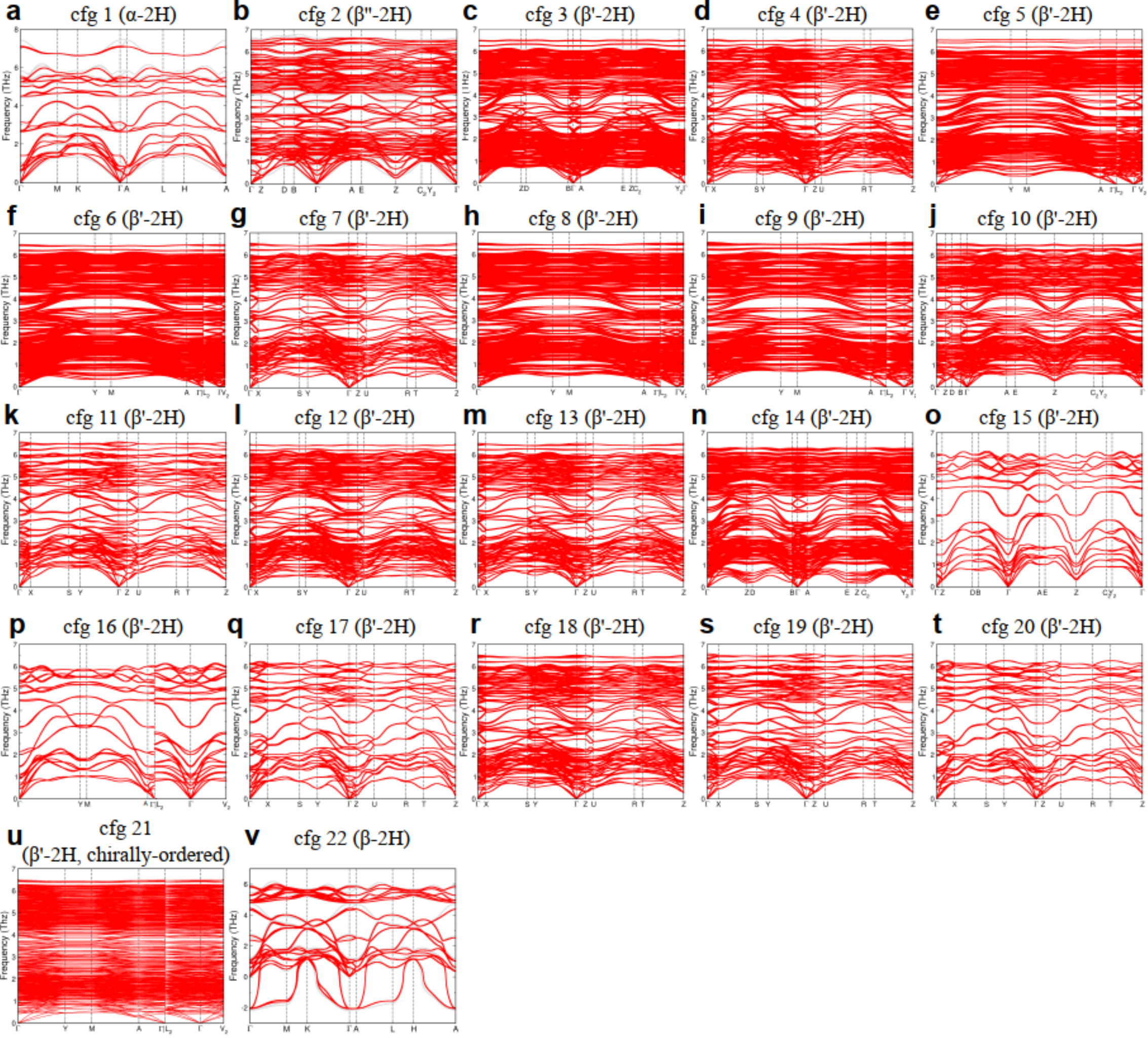}
		\caption{\textbf{| Phonon dispersions at 0 K for all identified polymorphs of vdWs In$_2$Se$_3$ with the 2H polytype.}
The phonon dispersions predicted by MLP are depicted in red lines, except for $\alpha$-2H (cfg1), $\beta''$-2H (cfg2), and $\beta$-2H (cfg22),
where DFT-calculated phonon dispersions are additionally shown in gray lines for comparison.
The corresponding crystal structures are presented in Supplementary Fig.~\ref{fig:FigS2}.
	}
	\label{fig:FigS4}
\end{figure*}

\newpage
\clearpage
\begin{figure*}[ht!]
	\centering
\includegraphics[width=0.95\textwidth]{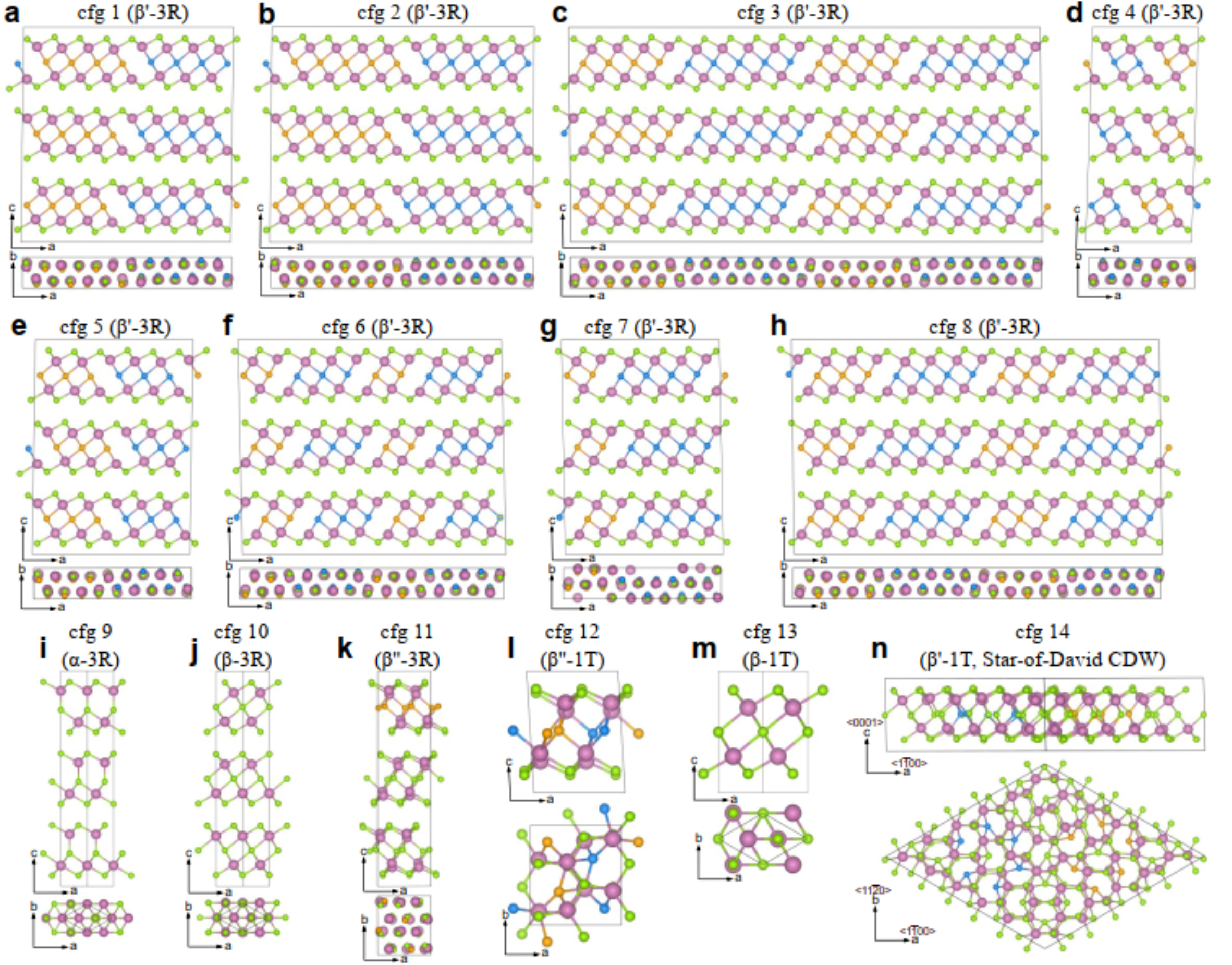}
		\caption{\textbf{| Identified polymorphs of vdWs In$_2$Se$_3$ in the 1T and 3R polytypes.}
Note that the large balls indicate the In atoms, whereas the small balls denote the Se atoms.
The orange and blue small balls in \textbf{a}-\textbf{h} represent the antiparallel displacements of central Se atoms along the $\langle11\bar{2}0\rangle$ direction,
while they are just used to guide the eye for the displacements of specified central Se atoms in  \textbf{k}, \textbf{l}, and  \textbf{n}.
	}
	\label{fig:FigS5}
\end{figure*}

\newpage
\clearpage
\begin{figure*}[ht!]
	\centering
\includegraphics[width=0.95\textwidth]{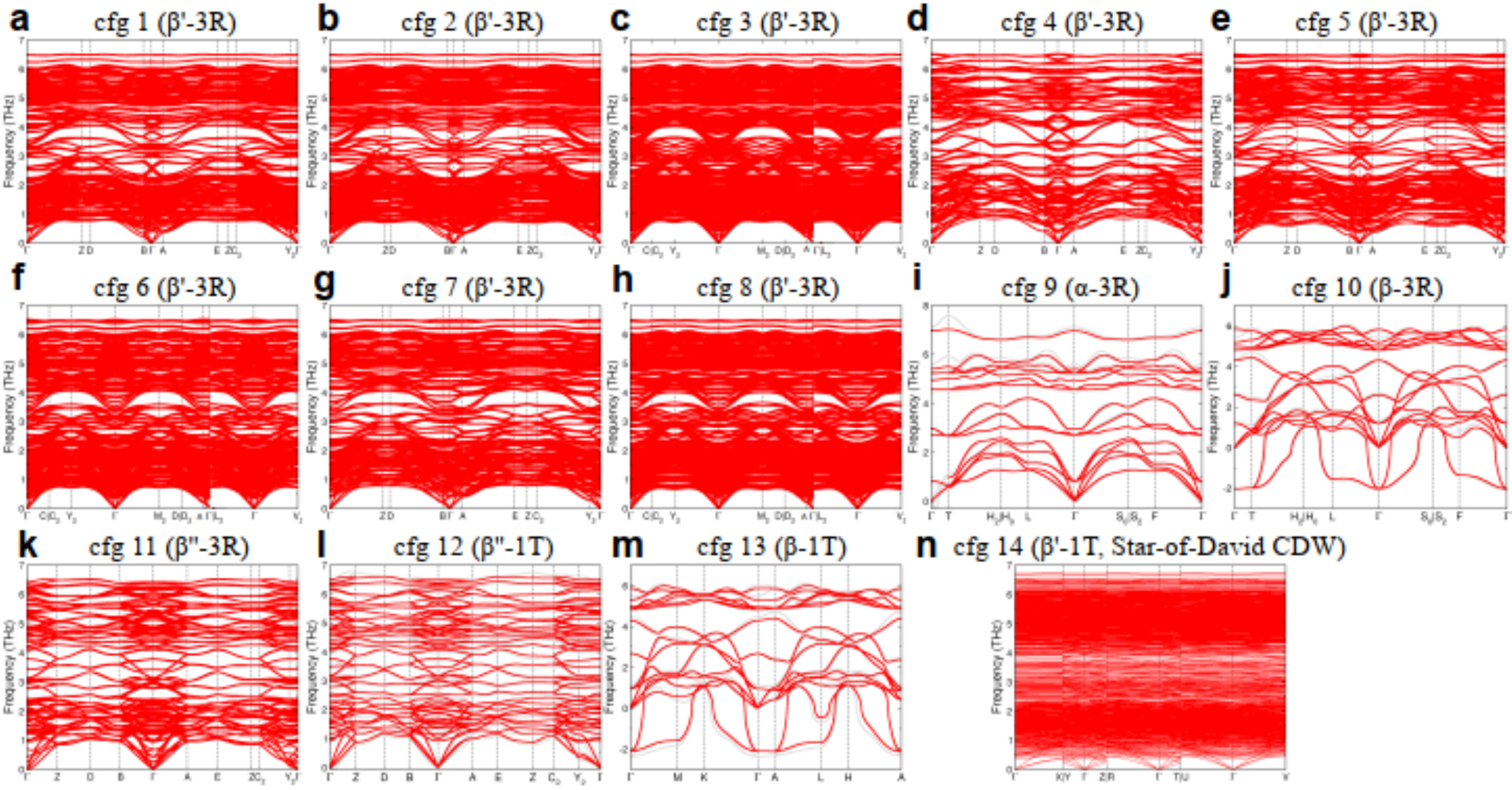}
		\caption{\textbf{| Phonon dispersions at 0 K for all identified polymorphs of vdWs In$_2$Se$_3$ with the 1T or 3R polytypes.}
The phonon dispersions predicted by MLP are depicted in red lines, except for $\alpha$-3R (cfg9), $\beta$-3R (cfg10), $\beta''$-1T (cfg12), and $\beta$-1T (cfg13),
where DFT-calculated phonon dispersions are additionally shown in gray lines for comparison.
The corresponding crystal structures are presented in Supplementary Fig.~\ref{fig:FigS5}.
	}
	\label{fig:FigS6}
\end{figure*}

\newpage
\clearpage
\begin{figure*}[ht!]
	\centering
\includegraphics[width=0.95\textwidth]{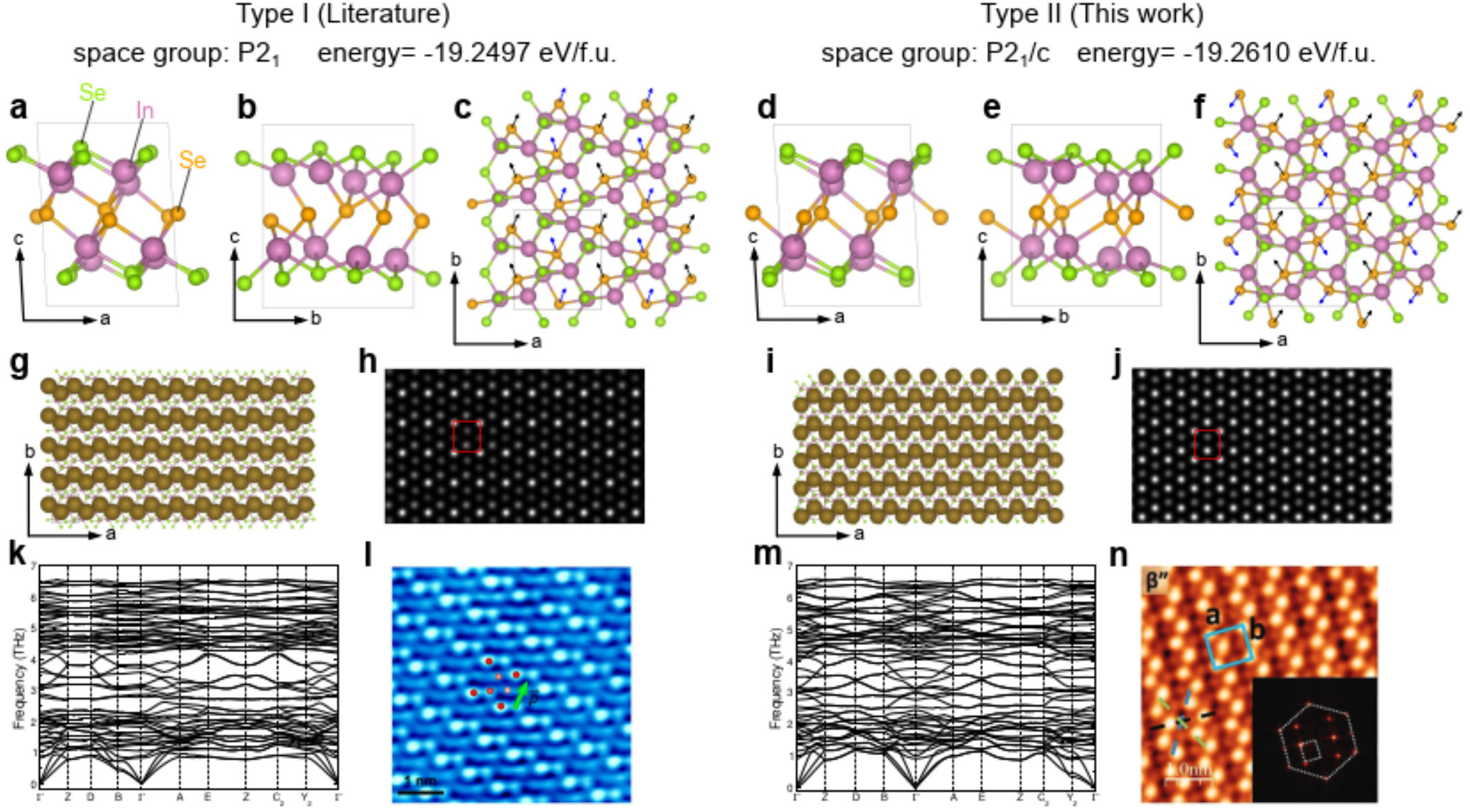}
		\caption{\textbf{| Identification of the crystal structure of the $\beta''$ phase.}
			\textbf{a}-\textbf{c} Different views of the crystal structure of the $\beta''$ phase proposed in
Refs.~\cite{ACSNano_Zhang2019,AdvancedScience_Chen2021,NC_beta1_beta2_electric_field_Zhang2024}  (referred to as "Type-I").
            \textbf{d}-\textbf{f} Different views of the crystal structure of the $\beta''$ phase identified in this work (referred to as "Type-II").
The space groups and DFT-calculated total energies are provided.
The central Se atoms are highlighted in the orange color.
The arrows in \textbf{c} and \textbf{f} denote the polarization directions associated with the central Se atoms.
            \textbf{g}, \textbf{i} Top view of the Type-I and Type-II structures, respectively.
            The Se atoms in the two topmost layers are enlarged to highlight the zig-zag striped morphology.
            \textbf{h}, \textbf{j} Simulated scanning tunneling microscopy (STM) images of the Type-I and Type-II structures, respectively.
            \textbf{k}, \textbf{m} MLP-predicted phonon dispersions of the Type-I and Type-II structures, respectively.
            \textbf{i}, \textbf{n} Experimental STM images reproduced from Ref.~\cite{NC_beta1_beta2_electric_field_Zhang2024} and Ref.~\cite{AdvancedScience_Chen2021}, respectively.
	}
	\label{fig:FigS7}
\end{figure*}

\newpage
\clearpage
\begin{figure*}[ht!]
	\centering
\includegraphics[width=0.9\textwidth]{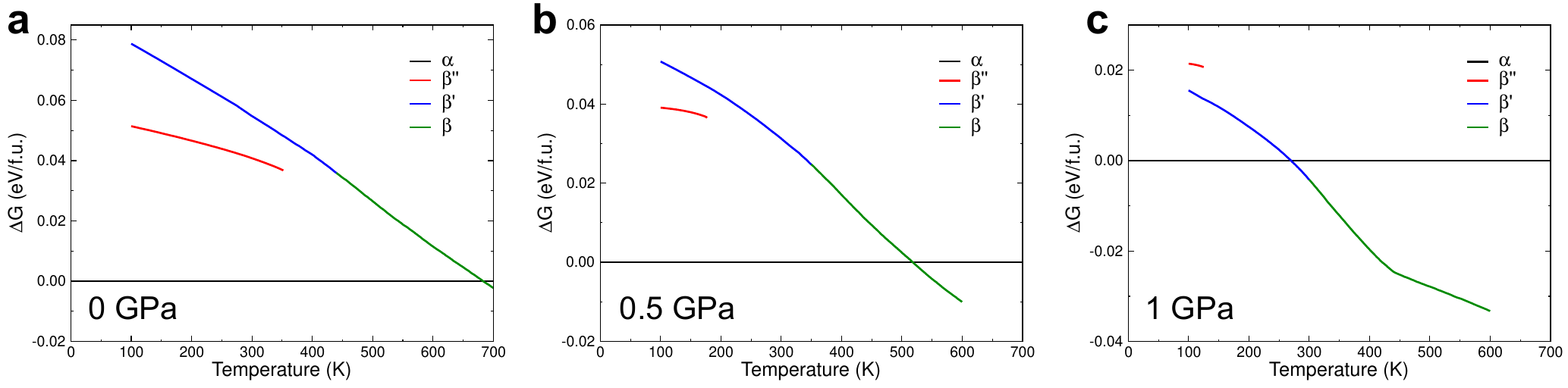}
		\caption{\textbf{| Gibbs free energies of $\alpha$-2H, $\beta$-2H, $\beta'$-2H, and $\beta''$-2H as a function of temperature at different pressures.}
\textbf{a} 0 GPa.
\textbf{b} 0.5 GPa.
\textbf{c} 1 GPa.
All Gibbs free energies are calculated relative to those of $\alpha$-2H.
	}
	\label{fig:FigS8}
\end{figure*}

\newpage
\clearpage
\begin{figure*}[ht!]
	\centering
\includegraphics[width=0.9\textwidth]{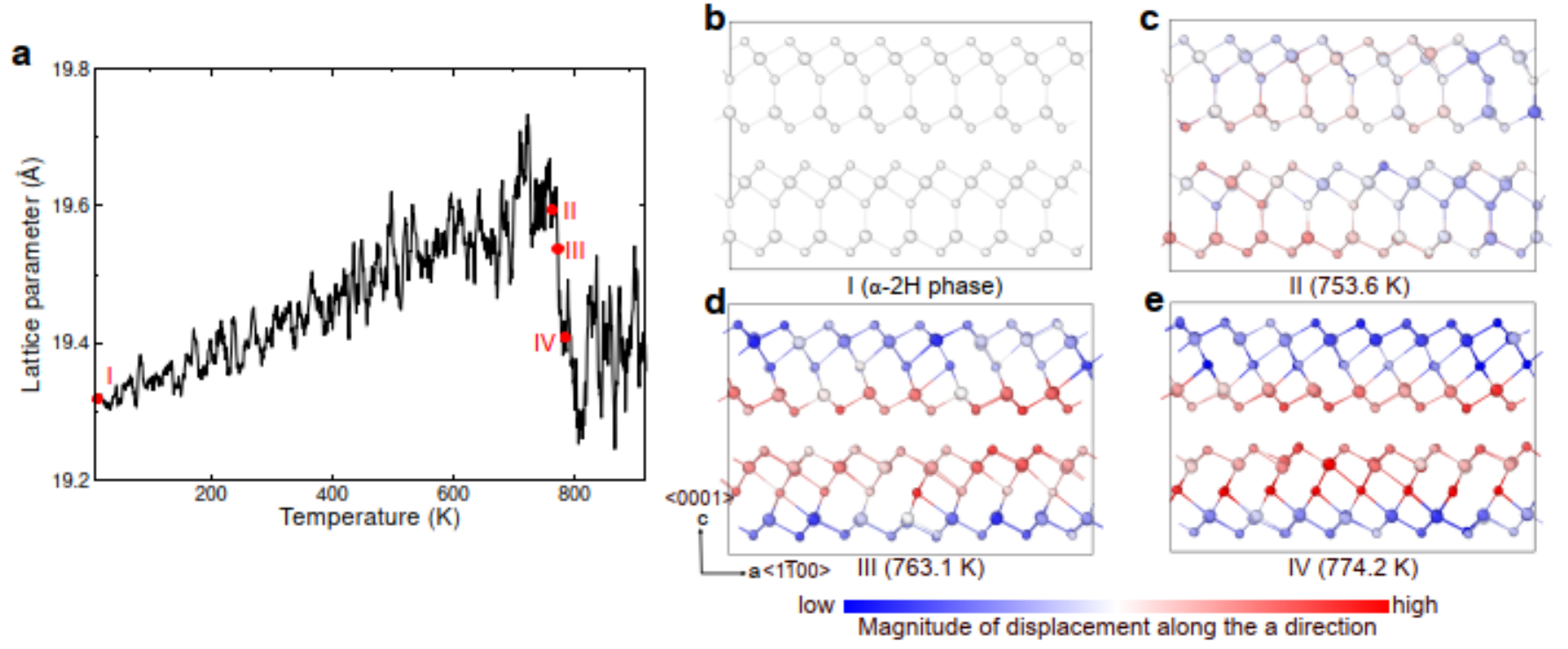}
		\caption{\textbf{| Temperature-induced $\alpha$-2H to $\beta$-2H  phase transition revealed by MD simulations.}
            \textbf{a} Evolution of the lattice parameter $c$ during heating  (heating rate: 0.053 K/ps) at 0 GPa.
The initial $\alpha$-2H supercell including 160 atoms is built by replicating the $\alpha$-2H unit cell using the transformation matrix [4~4~0; $-$2~2~0; 0 0 1].
   			\textbf{b}-\textbf{e} MD snapshots corresponding to the simulation times marked by red dots in \textbf{a}.
            The large and small balls indicate the In and Se atoms, respectively.
            The color represents the magnitude of displacement along the $a$ direction relative to the $\alpha$-2H phase.
	}
	\label{fig:FigS9}
\end{figure*}

\newpage
\clearpage
\begin{figure*}[ht!]
	\centering
\includegraphics[width=0.95\textwidth]{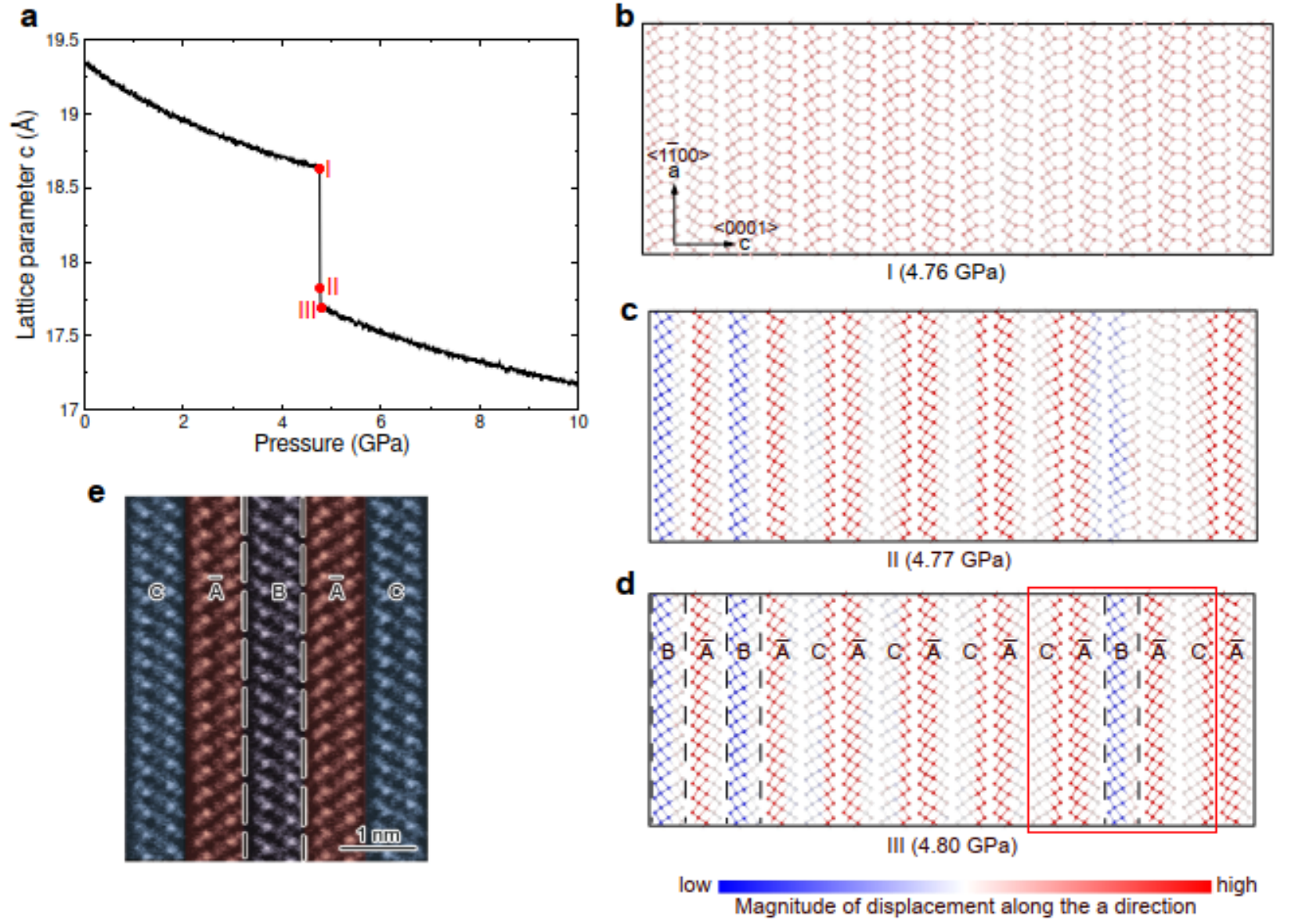}
		\caption{\textbf{| Pressure-induced phase transition from $\alpha$-2H to $\beta$-2H revealed by MD simulations.}
			\textbf{a} Evolution of the lattice parameter $c$ during pressurization (pressurization  rate: 0.67 GPa/ns) at 100 K.
            The initial $\alpha$-2H supercell including 2,560 atoms is constructed by replicating the $\alpha$-2H unit cell using the transformation matrix [8~8~0; $-$2~2~0; 0 0 8].
   			\textbf{b}-\textbf{d} MD snapshots corresponding to the simulation times marked by red dots in \textbf{a}.
            The color represents the magnitude of displacement along the $a$ direction relative to the $\alpha$-2H phase.
            \textbf{e} Atomic-scale high-angle annular dark-field scanning transmission electron microscopy image of $\beta$-2H In$_2$Se$_3$
            reproduced from our previous experiment~\cite{Zhibo_SF_2024}.
            The replacement-type stacking faults are highlighted by the dashed lines in \textbf{d} and \textbf{e}.
            The region marked by the rectangle in \textbf{d} matches excellently with the experimental image  in \textbf{e}.
            The notations are taken from Ref.~\cite{Zhibo_SF_2024}. Specifically, the capital letter represents the stacking sequence of the central Se-atom layer.
            For instance, A, B, and C represent the stacking sequences of $cabca$, $cbacb$, and $bacba$ for the quintuple layer, respectively.
            The $\bar{\rm A}$ denotes the 180$^\circ$ rotation of the A quintuple layer.
	}
	\label{fig:FigS10}
\end{figure*}

\newpage
\clearpage
\begin{figure*}[ht!]
	\centering
\includegraphics[width=0.95\textwidth]{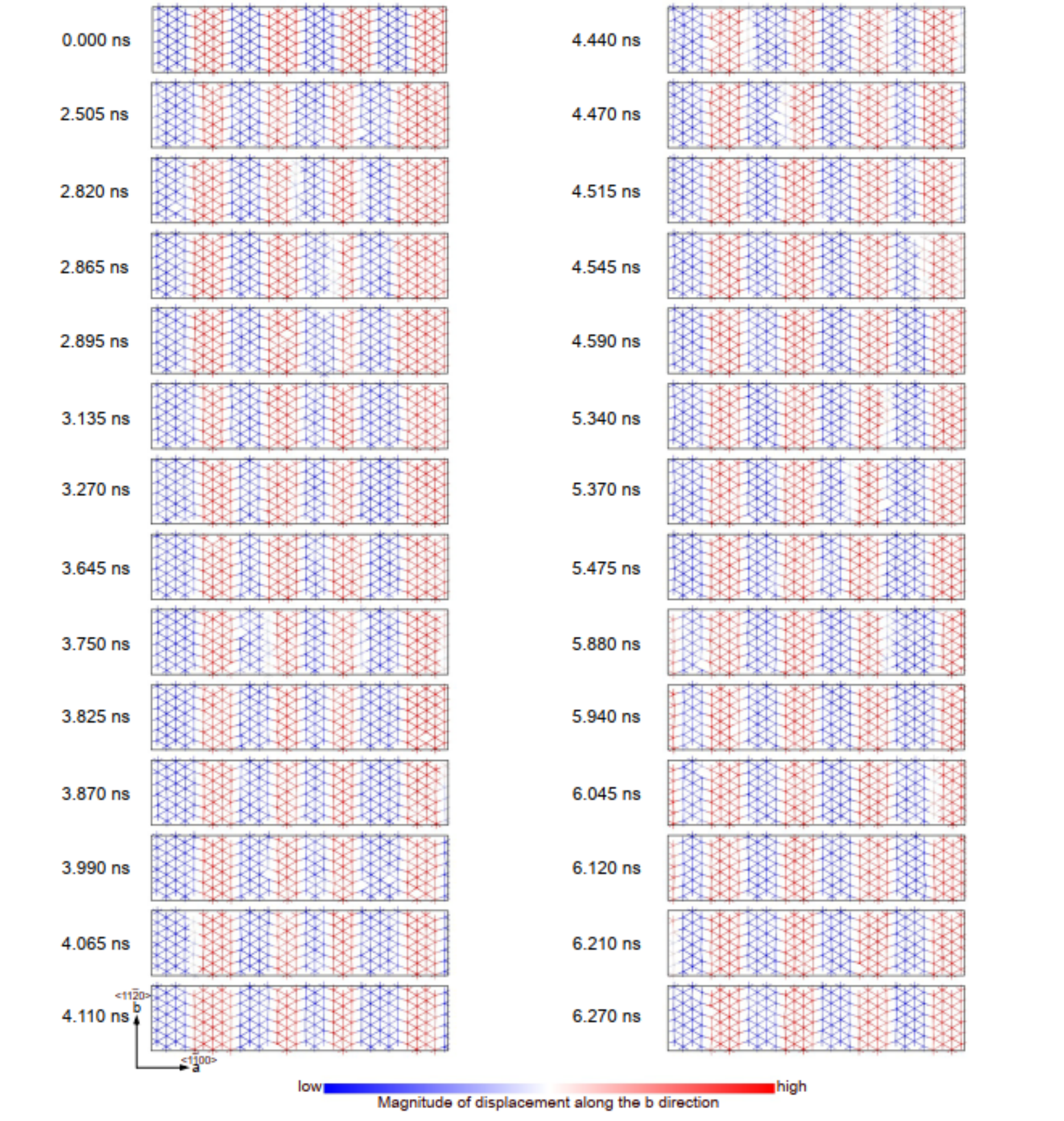}
		\caption{\textbf{| Dynamical motion of domain walls in the $\beta'$-2H polymorph.}
Time-resolved molecular dynamics snapshots at 300 K and 0 GPa illustrate the evolution of domain walls.
 The initial $\beta'$-2H supercell including 1,920 atoms is built by replicating the unit cell of the 4$d$-4$\bar{d}$/4$d$-4$\bar{d}$ variant in a $4\times6\times1$ configuration.
 For clarity, only the central Se atoms within one quintuple layer are displayed, highlighting the domain walls.
 The color scale indicates the magnitude of atomic displacement along the $b$ direction, relative to the ideal undistorted $\beta$-2H phase.
	}
	\label{fig:FigS11}
\end{figure*}

\newpage
\clearpage
\begin{figure*}[ht!]
	\centering
\includegraphics[width=0.7\textwidth]{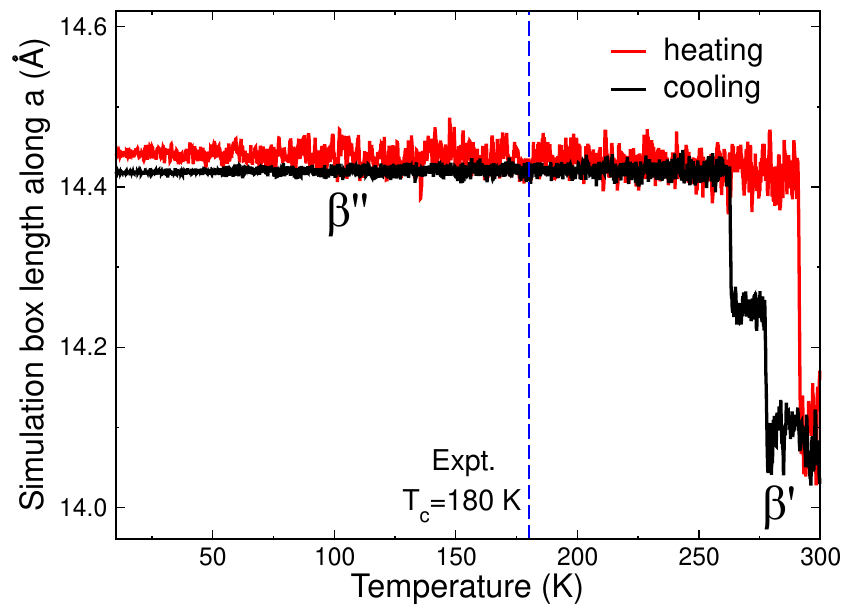}
		\caption{\textbf{| Temperature-induced reversible phase transitions between $\beta'$-2H and $\beta''$-2H revealed by MD simulations.}
           Evolution of the simulation box length along $a$-axis during heating (heating rate: 0.039 K/ps) or cooling (cooling rate: 0.39 K/ps) at 0 GPa.
           Note that the high cooling rate is necessary for obtaining the $\beta''$-2H phase, as it facilitates the quenching of the $\beta'$-2H phase.
           The initial $\beta'$-2H supercell is constructed by replicating the unit cell of the 4$d$-4$\bar{d}$/4$d$-4$\bar{d}$ variant  in a $1\times4\times1$ configuration.
           The initial $\beta''$-2H supercell is built by replicating the unit cell of the $\beta''$-2H in a $2\times2\times1$ configuration.
           Both supercells consist of 160 atoms.
            Note that the two abrupt jumps observed during the cooling process arise from layer-by-layer transformations.
            The experimentally determined phase transition temperature is around 180 K~\cite{ACSNano_Zhang2019}, as indicated by the blue dashed line.
	}
	\label{fig:FigS12}
\end{figure*}

\newpage
\clearpage
\begin{figure*}[ht!]
	\centering
\includegraphics[width=0.95\textwidth]{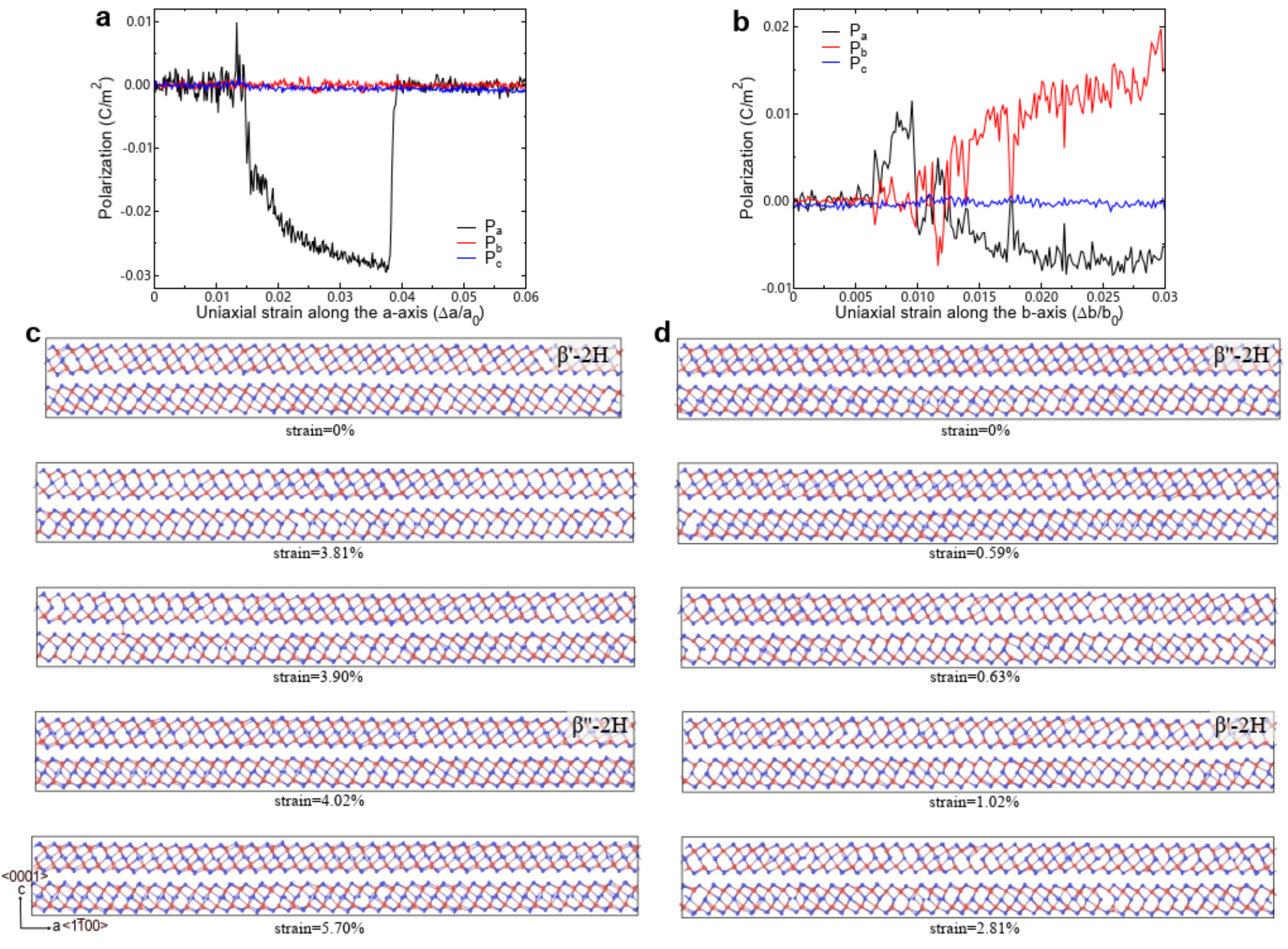}
		\caption{\textbf{| Strain-induced polarization modulations.}
			\textbf{a} Evolution of polarizations under uniaxial strain along the $a$-axis at 300 K and 0 GPa, starting from the $\beta'$-2H phase.
			\textbf{b} Evolution of polarizations under uniaxial strain along the $b$-axis at 300 K and 0 GPa, starting from the $\beta''$-2H phase.
			\textbf{c} MD snapshots under selected strains in \textbf{a}.
			\textbf{d} MD snapshots under selected strains in \textbf{b}.
            Note that these two MD simulations correspond to those presented in Fig.~6 of the main text. The blue and red balls indicate Se and In atoms, respectively.
	}
	\label{fig:FigS13}
\end{figure*}

\newpage
\clearpage
\begin{figure*}[ht!]
	\centering
\includegraphics[width=0.95\textwidth]{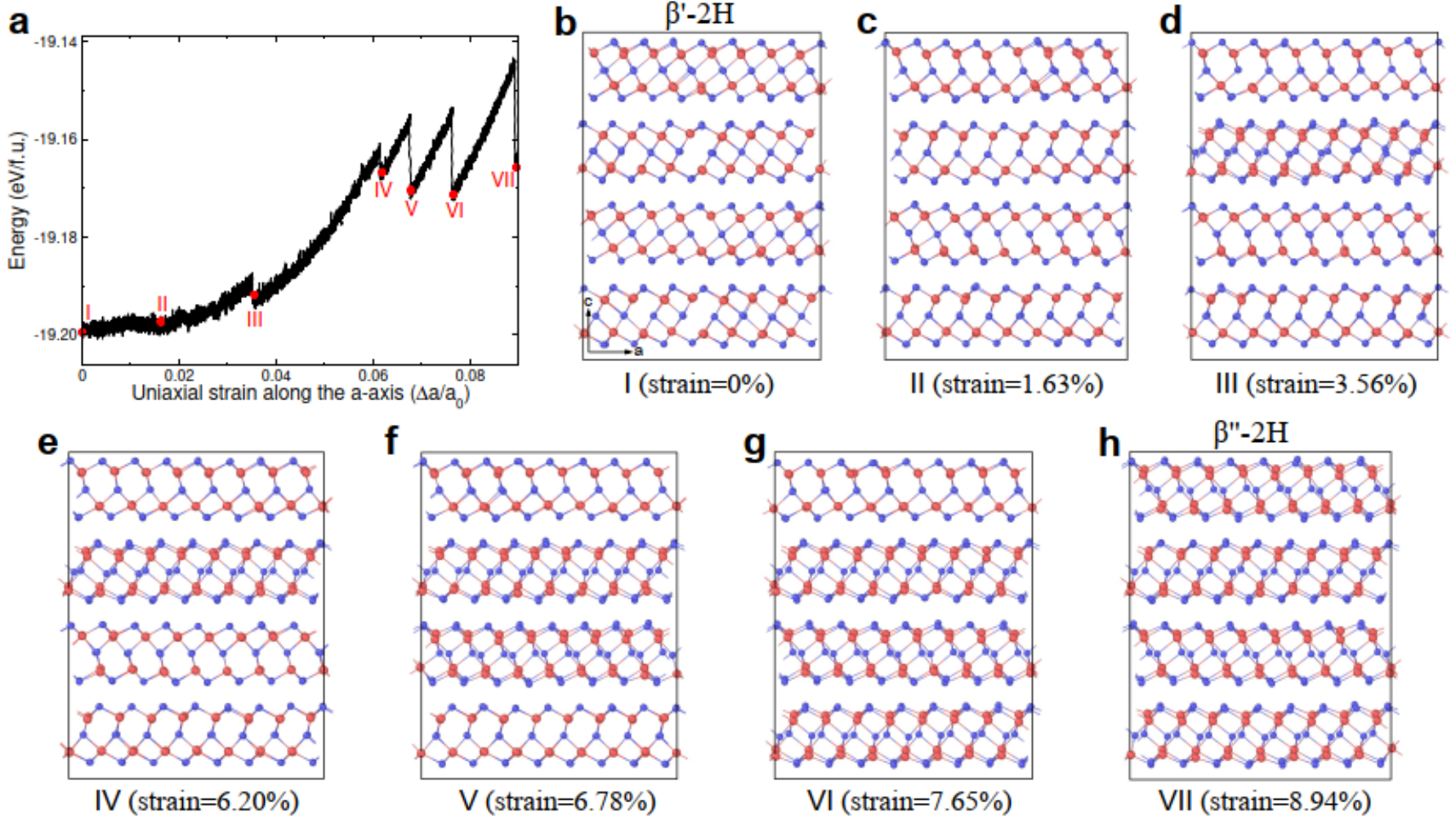}
		\caption{\textbf{| Strain-induced layer-by-layer phase transition from $\beta'$-2H to $\beta''$-2H revealed by MD simulations.}
			\textbf{a} Evolution of the potential energy under uniaxial strain along the $a$-axis (strain rate: 0.01 ns$^{-1}$) at 100 K and 0 GPa, starting from the $\beta'$-2H phase.
            The initial structures for the MD simulations are based on a 320-atom supercell, constructed by $1\times2\times2$ replications of the 4$d$-4$\bar{d}$/4$d$-4$\bar{d}$ variant of the $\beta'$-2H phase.
			\textbf{b}-\textbf{h} MD snapshots corresponding to the strain values marked by the red dots in \textbf{a},
           highlighting the strain-induced layer-by-layer transformation from $\beta'$-2H to $\beta''$-2H.
	}
	\label{fig:FigS14}
\end{figure*}

\newpage
\clearpage
\section*{Supplementary Tables}
\vspace{2mm}

\begin{table}[ht!]
	\caption{Lattice parameters predicted by DFT and MLP for identified polymorphs with the 2H polytype, compared with available experimental data.
Crystal structures of these polymorphs are shown in Supplementary Fig.~\ref{fig:FigS2}.
}
	\begin{tabular}{p{1cm}p{2cm}p{2.5cm}p{1.6cm}p{1.8cm}p{1.8cm}p{1.8cm}p{1.8cm}p{2cm}}
		\toprule[0.4mm]
		\multirow{2}{*}{Polytype}    & \multirow{2}{*}{Polymorph} & \multirow{2}{*}{Space   group} & \multirow{2}{*}{}                             & \multicolumn{4}{c}{Lattice parameters}                                                                                                                                                                                                                             \\ \cline{5-8}
		&           &           &           & $a$ (~\AA)            & $b$ (~\AA)            & $c$ (~\AA)           &$\alpha$ /$\beta$ /$\gamma$         \\ \hline
		\multirow{24}{*}{2H} & cfg1                   & P6$_3$mc   (186)                  & \begin{tabular}[c]{@{}l@{}}DFT\\    MLP   \\ Expt.\cite{ChemicalReviews_Tan2023} \end{tabular} & \begin{tabular}[c]{@{}l@{}}4.071\\   4.073 \\ 4.02 \end{tabular}   & \begin{tabular}[c]{@{}l@{}}4.071\\  4.073   \\ 4.02 \end{tabular} & \begin{tabular}[c]{@{}l@{}}19.368\\    19.320   \\ 19.22 \end{tabular} & \begin{tabular}[c]{@{}l@{}}$\gamma$=120$^{\rm o}$\\    $\gamma$=120$^{\rm o}$   \\ $\gamma$=120$^{\rm o}$ \end{tabular}       \\
\cline{2-8}
		& cfg2    & P2$_1$ (4)    & \begin{tabular}[c]{@{}l@{}}DFT\\   MLP\end{tabular} & \begin{tabular}[c]{@{}l@{}}7.184\\  7.202\end{tabular}   & \begin{tabular}[c]{@{}l@{}}7.957\\  7.937\end{tabular} & \begin{tabular}[c]{@{}l@{}}19.055\\  19.088\end{tabular} & \begin{tabular}[c]{@{}l@{}}$\beta$=90.09$^{\rm o}$\\  $\beta$=90.14$^{\rm o}$\end{tabular}   \\
\cline{2-8}
		& cfg3                   & Pna2$_1$   (33)                   & \begin{tabular}[c]{@{}l@{}}DFT\\   MLP\end{tabular} & \begin{tabular}[c]{@{}l@{}}34.368\\  34.447\end{tabular} & \begin{tabular}[c]{@{}l@{}}4.106\\   4.099\end{tabular} & \begin{tabular}[c]{@{}l@{}}18.939\\   18.928\end{tabular} & \begin{tabular}[c]{@{}l@{}}$\beta$=90$^{\rm o}$\\  $\beta$=90$^{\rm o}$\end{tabular}         \\
\cline{2-8}
		& cfg4                   & Pca2$_1$   (29)                   & \begin{tabular}[c]{@{}l@{}}DFT\\   MLP\end{tabular} & \begin{tabular}[c]{@{}l@{}}27.625\\   27.659\end{tabular} & \begin{tabular}[c]{@{}l@{}}4.095\\  4.094\end{tabular} & \begin{tabular}[c]{@{}l@{}}18.948\\   18.965\end{tabular} & \begin{tabular}[c]{@{}l@{}}$\beta$=90$^{\rm o}$\\   $\beta$=90$^{\rm o}$\end{tabular}         \\
\cline{2-8}
		& cfg5                   & Cc (9)                         & \begin{tabular}[c]{@{}l@{}}DFT\\   MLP\end{tabular} & \begin{tabular}[c]{@{}l@{}}35.313\\  35.186\end{tabular} & \begin{tabular}[c]{@{}l@{}}4.011\\   4.018\end{tabular} & \begin{tabular}[c]{@{}l@{}}18.959\\   18.979\end{tabular} & \begin{tabular}[c]{@{}l@{}}$\beta$=90.36$^{\rm o}$\\   $\beta$=90.24$^{\rm o}$\end{tabular}   \\
\cline{2-8}
		& cfg6                   & Cc   (9)                       & \begin{tabular}[c]{@{}l@{}}DFT\\   MLP\end{tabular} & \begin{tabular}[c]{@{}l@{}}62.047\\   62.099\end{tabular} & \begin{tabular}[c]{@{}l@{}}4.098\\   4.097\end{tabular} & \begin{tabular}[c]{@{}l@{}}18.940\\   18.944\end{tabular} & \begin{tabular}[c]{@{}l@{}}$\beta$=89.97$^{\rm o}$\\   $\beta$=90.06$^{\rm o}$\end{tabular}   \\
\cline{2-8}
		& cfg7                   & Pna2$_1$ (33)                     & \begin{tabular}[c]{@{}l@{}}DFT\\   MLP\end{tabular} & \begin{tabular}[c]{@{}l@{}}20.762\\   20.767\end{tabular} & \begin{tabular}[c]{@{}l@{}}4.094\\   4.095\end{tabular} & \begin{tabular}[c]{@{}l@{}}18.925\\   18.964\end{tabular} & \begin{tabular}[c]{@{}l@{}}$\beta$=90$^{\rm o}$\\   $\beta$=90$^{\rm o}$\end{tabular}         \\
\cline{2-8}
		& cfg8                   & Cc   (9)                       & \begin{tabular}[c]{@{}l@{}}DFT\\   MLP\end{tabular} & \begin{tabular}[c]{@{}l@{}}48.413\\    48.414\end{tabular} & \begin{tabular}[c]{@{}l@{}}4.093\\   4.095\end{tabular} & \begin{tabular}[c]{@{}l@{}}18.940\\  18.959\end{tabular} & \begin{tabular}[c]{@{}l@{}}$\beta$=89.97$^{\rm o}$   \\ $\beta$=90.07$^{\rm o}$\end{tabular}   \\
\cline{2-8}
		& cfg9                   & Cc   (9)                       & \begin{tabular}[c]{@{}l@{}}DFT\\    MLP\end{tabular} & \begin{tabular}[c]{@{}l@{}}34.726\\    34.748\end{tabular} & \begin{tabular}[c]{@{}l@{}}4.087\\   4.090\end{tabular} & \begin{tabular}[c]{@{}l@{}}18.943\\   18.985\end{tabular} & \begin{tabular}[c]{@{}l@{}}$\beta$=90.16$^{\rm o}$\\  $\beta$=90.05$^{\rm o}$\end{tabular}   \\
\cline{2-8}
		& cfg10                  & P2$_1$   (4)                      & \begin{tabular}[c]{@{}l@{}}DFT\\   MLP\end{tabular} & \begin{tabular}[c]{@{}l@{}}24.289\\   24.349\end{tabular} & \begin{tabular}[c]{@{}l@{}}4.093\\   4.091\end{tabular} & \begin{tabular}[c]{@{}l@{}}18.944\\   19.001\end{tabular} & \begin{tabular}[c]{@{}l@{}}$\beta$=94.71$^{\rm o}$\\   $\beta$=94.79$^{\rm o}$\end{tabular}   \\
\cline{2-8}
		& cfg11                  & Pca2$_1$   (29)                   & \begin{tabular}[c]{@{}l@{}}DFT\\    MLP\end{tabular} & \begin{tabular}[c]{@{}l@{}}13.914\\    13.977\end{tabular} & \begin{tabular}[c]{@{}l@{}}4.087\\    4.084\end{tabular} & \begin{tabular}[c]{@{}l@{}}18.907\\   19.003\end{tabular} & \begin{tabular}[c]{@{}l@{}}$\beta$=90$^{\rm o}$\\   $\beta$=90$^{\rm o}$\end{tabular}         \\
\cline{2-8}
		& cfg12                  & Pna2$_1$   (33)                   & \begin{tabular}[c]{@{}l@{}}DFT\\   MLP\end{tabular} & \begin{tabular}[c]{@{}l@{}}27.651\\  27.650\end{tabular} & \begin{tabular}[c]{@{}l@{}}4.089\\    4.092\end{tabular} & \begin{tabular}[c]{@{}l@{}}19.031\\   19.029\end{tabular} & \begin{tabular}[c]{@{}l@{}}$\beta$=90$^{\rm o}$\\   $\beta$=90$^{\rm o}$\end{tabular}        \\
\cline{2-8}
		& cfg13                  & Pca2$_1$   (29)                   & \begin{tabular}[c]{@{}l@{}}DFT\\    MLP\end{tabular} & \begin{tabular}[c]{@{}l@{}}20.753   \\ 20.773\end{tabular} & \begin{tabular}[c]{@{}l@{}}4.093\\   4.092\end{tabular} & \begin{tabular}[c]{@{}l@{}}18.983\\   19.044\end{tabular} & \begin{tabular}[c]{@{}l@{}}$\beta$=90$^{\rm o}$   \\ $\beta$=90$^{\rm o}$\end{tabular}         \\
\cline{2-8}
		& cfg14                  & Pc   (7)                       & \begin{tabular}[c]{@{}l@{}}DFT\\   MLP\end{tabular} & \begin{tabular}[c]{@{}l@{}}28.309   \\ 28.258\end{tabular} & \begin{tabular}[c]{@{}l@{}}3.986\\   3.990\end{tabular} & \begin{tabular}[c]{@{}l@{}}18.946\\   18.907\end{tabular} & \begin{tabular}[c]{@{}l@{}}$\beta$=90.53$^{\rm o}$   \\ $\beta$=90.39$^{\rm o}$\end{tabular}     \\
\cline{2-8}
		& cfg15                  & P2$_1$   (4)                      & \begin{tabular}[c]{@{}l@{}}DFT\\   MLP\end{tabular} & \begin{tabular}[c]{@{}l@{}}3.977   \\ 3.978\end{tabular}   & \begin{tabular}[c]{@{}l@{}}3.990\\   3.990\end{tabular} & \begin{tabular}[c]{@{}l@{}}18.994\\   18.920\end{tabular} & \begin{tabular}[c]{@{}l@{}}$\beta$=117.93$^{\rm o}$   \\ $\beta$=118.01$^{\rm o}$\end{tabular} \\
\cline{2-8}
		& cfg16                  & Cc   (9)                       & \begin{tabular}[c]{@{}l@{}}DFT\\   MLP\end{tabular} & \begin{tabular}[c]{@{}l@{}}6.819   \\ 6.830\end{tabular}   & \begin{tabular}[c]{@{}l@{}}4.108\\   4.103\end{tabular} & \begin{tabular}[c]{@{}l@{}}18.991\\  18.938\end{tabular} & \begin{tabular}[c]{@{}l@{}}$\beta$=90.70$^{\rm o}$\\  $\beta$=90.48$^{\rm o}$\end{tabular}   \\
\cline{2-8}
		& cfg17                  & Pna2$_1$   (33)                   & \begin{tabular}[c]{@{}l@{}}DFT\\   MLP\end{tabular} & \begin{tabular}[c]{@{}l@{}}7.074  \\ 7.066\end{tabular}   & \begin{tabular}[c]{@{}l@{}}3.984\\   3.988\end{tabular} & \begin{tabular}[c]{@{}l@{}}18.941\\   18.922\end{tabular} & \begin{tabular}[c]{@{}l@{}}$\beta$=90$^{\rm o}$  \\ $\beta$=90$^{\rm o}$\end{tabular}         \\
\cline{2-8}
		& cfg18                  & Pca2$_1$   (29)                   & \begin{tabular}[c]{@{}l@{}}DFT\\   MLP\end{tabular} & \begin{tabular}[c]{@{}l@{}}20.788  \\ 20.765\end{tabular} & \begin{tabular}[c]{@{}l@{}}4.084\\   4.088\end{tabular} & \begin{tabular}[c]{@{}l@{}}19.031\\   19.053\end{tabular} & \begin{tabular}[c]{@{}l@{}}$\beta$=90$^{\rm o}$   \\ $\beta$=90$^{\rm o}$\end{tabular}         \\
\cline{2-8}
		& cfg19                  & Pca2$_1$   (29)                   & \begin{tabular}[c]{@{}l@{}}DFT\\   MLP\end{tabular} & \begin{tabular}[c]{@{}l@{}}13.956  \\ 13.968\end{tabular} & \begin{tabular}[c]{@{}l@{}}4.073\\    4.076\end{tabular} & \begin{tabular}[c]{@{}l@{}}19.051\\   19.082\end{tabular} & \begin{tabular}[c]{@{}l@{}}$\beta$=90$^{\rm o}$  \\ $\beta$=90$^{\rm o}$\end{tabular}  \\
\cline{2-8}
		& cfg20                  & Pca2$_1$   (29)                   & \begin{tabular}[c]{@{}l@{}}DFT\\  MLP\end{tabular}   & \begin{tabular}[c]{@{}l@{}}7.071\\  7.066\end{tabular}   & \begin{tabular}[c]{@{}l@{}}3.972\\    3.980\end{tabular}    & \begin{tabular}[c]{@{}l@{}}19.012\\    18.961\end{tabular}     & \begin{tabular}[c]{@{}l@{}}$\beta$=90$^{\rm o}$\\   $\beta$=90$^{\rm o}$\end{tabular}  \\
\cline{2-8}
		& cfg21                  & Cc   (9)                       & \begin{tabular}[c]{@{}l@{}}DFT\\ \\ \\  MLP\end{tabular} & \begin{tabular}[c]{@{}l@{}}\\12.187\\ \\ \\  12.205 \\ \\ \end{tabular} & \begin{tabular}[c]{@{}l@{}}  12.187 \\ \\ \\  12.205\end{tabular} & \begin{tabular}[c]{@{}l@{}}18.931\\ \\ \\    18.922\end{tabular} & \begin{tabular}[c]{@{}l@{}}$\alpha$=90.14$^{\rm o}$\\    $\beta$=90.14$^{\rm o}$\\    $\gamma$=120.74$^{\rm o}$\\   $\alpha$=90.11$^{\rm o}$ \\ $\beta$=90.11$^{\rm o}$ \\ $\gamma$=120.84$^{\rm o}$\end{tabular} \\
\cline{2-8}
		& cfg22                  & P6$_3$mc   (186)                  & \begin{tabular}[c]{@{}l@{}}DFT\\    MLP  \\ Expt.\cite{ChemicalReviews_Tan2023} \end{tabular}                     & \begin{tabular}[c]{@{}l@{}}3.990\\   3.990  \\ 4.01 \end{tabular}       & \begin{tabular}[c]{@{}l@{}}3.990\\    3.990  \\4.01 \end{tabular}     & \begin{tabular}[c]{@{}l@{}}18.723\\   18.670  \\19.48 \end{tabular}        & \begin{tabular}[c]{@{}l@{}}$\gamma$=120$^{\rm o}$\\    $\gamma$=120$^{\rm o}$  \\ $\gamma$=120$^{\rm o}$ \end{tabular}\\
\toprule[0.4mm]
	\end{tabular}
	\label{tab:TableS1}
\end{table}


\newpage
\clearpage
\begin{table}[htbp]
	\caption{Predicted periodic lengths of the $\beta'$ superstructures using MLP, compared with experimental results.}
	\centering
	\begin{tabular}{cccc}
		\toprule[0.4mm]
Modulated superstructures & \multicolumn{2}{c}{Periodic length (Predictions) (\AA)} & Periodic length (\AA) \\
 \cline{2-3}
 &  $\beta'$-2H & $\beta'$-3R &  (Experiments) \\
 \hline
		2$d$-2$\bar{d}$/2$d$-2$\bar{d}$ & 13.98 & 13.99 & 14~\cite{van1975,Lin2013_JACS} \\
		2$d$-3$\bar{d}$/2$d$-3$\bar{d}$ & 17.37 & 17.40 & 17.5~\cite{van1975} \\
		3$d$-3$\bar{d}$/3$d$-3$\bar{d}$ & 20.77 & 20.81 & 21~\cite{van1975} \\
		3$d$-4$\bar{d}$/3$d$-4$\bar{d}$ & 24.21 & 24.24 & --- \\
		4$d$-4$\bar{d}$/4$d$-4$\bar{d}$ & 27.66 & 27.67 & 28.0~\cite{van1975,PRL_AFE2020,NC_ferroelasticity_Xu2021} \\
		4$d$-5$\bar{d}$/4$d$-5$\bar{d}$ & 31.05 & 31.09 & 31.5~\cite{van1975,PRL_AFE2020} \\
		5$d$-5$\bar{d}$/5$d$-5$\bar{d}$ & 34.45 & 34.51 & --- \\
		\toprule[0.4mm]
	\end{tabular}
	\label{tab:TableS2}
\end{table}

\begin{table}[ht!]
    \caption{Lattice parameters predicted by DFT and MLP for identified polymorphs with the 3R or 1T polytypes, compared with available experimental data.
Crystal structures of these polymorphs are shown in Supplementary Fig.~\ref{fig:FigS5}.}
	\begin{tabular}{p{1cm}p{2cm}p{2.5cm}p{1.6cm}p{1.8cm}p{1.8cm}p{1.8cm}p{1.8cm}p{2cm}}
		\toprule[0.4mm]
		\multirow{2}{*}{Polytype}    & \multirow{2}{*}{Polymorph} & \multirow{2}{*}{Space   group} & \multirow{2}{*}{}                             & \multicolumn{4}{c}{Lattice parameters}                                                                                                                                                                                                                                                                                                                                  \\ \cline{5-8}
		&      &      &       & a (~\AA)      & b (~\AA)       & c (~\AA)      &$\alpha$ /$\beta$ /$\gamma$         \\ \hline
		\multirow{24}{*}{3R} & cfg1                   & P2$_1$/c   (14)                   & \begin{tabular}[c]{@{}l@{}}DFT   \\ MLP\end{tabular}                      & \begin{tabular}[c]{@{}l@{}}27.641    \\ 27.673\end{tabular}                & \begin{tabular}[c]{@{}l@{}}4.093   \\ 4.091\end{tabular}                  & \begin{tabular}[c]{@{}l@{}}28.357  \\ 28.347\end{tabular}               & \begin{tabular}[c]{@{}l@{}} $\beta$=90.99$^{\rm o}$   \\ $\beta$=90.75$^{\rm o}$\end{tabular}                                                         \\
\cline{2-8}
		& cfg2                   & P2/c (13)                      & \begin{tabular}[c]{@{}l@{}}DFT   \\ MLP\end{tabular}                      & \begin{tabular}[c]{@{}l@{}}34.486  \\ 34.506\end{tabular}                & \begin{tabular}[c]{@{}l@{}}4.095   \\ 4.093\end{tabular}                  & \begin{tabular}[c]{@{}l@{}}28.351   \\ 28.326\end{tabular}               & \begin{tabular}[c]{@{}l@{}}$\beta$=90.99$^{\rm o}$   \\ $\beta$=90.56$^{\rm o}$\end{tabular}                                                         \\
\cline{2-8}
		& cfg3                   & C2 (5)                         & \begin{tabular}[c]{@{}l@{}}DFT\\    MLP\end{tabular}                      & \begin{tabular}[c]{@{}l@{}}62.190  \\ 62.180\end{tabular}                & \begin{tabular}[c]{@{}l@{}}4.091\\    4.092\end{tabular}                  & \begin{tabular}[c]{@{}l@{}}28.371 \\ 28.336\end{tabular}               & \begin{tabular}[c]{@{}l@{}}$\beta$=91.06$^{\rm o}$   \\ $\beta$=90.64$^{\rm o}$\end{tabular}                                                         \\
\cline{2-8}
		& cfg4                   & P2$_1$/c   (14)                   & \begin{tabular}[c]{@{}l@{}}DFT\\   MLP\end{tabular}                      & \begin{tabular}[c]{@{}l@{}}13.912   \\ 13.986\end{tabular}                & \begin{tabular}[c]{@{}l@{}}4.083\\   4.079\end{tabular}                  & \begin{tabular}[c]{@{}l@{}}28.377   \\ 28.433\end{tabular}               & \begin{tabular}[c]{@{}l@{}}$\beta$=91.02$^{\rm o}$\\    $\beta$=91.79$^{\rm o}$\end{tabular}                                                         \\
\cline{2-8}
		& cfg5                   & P2/c   (13)                    & \begin{tabular}[c]{@{}l@{}}DFT   \\ MLP\end{tabular}                      & \begin{tabular}[c]{@{}l@{}}20.807   \\ 20.806\end{tabular}                & \begin{tabular}[c]{@{}l@{}}4.088  \\ 4.089\end{tabular}                  & \begin{tabular}[c]{@{}l@{}}28.373  \\ 28.359\end{tabular}               & \begin{tabular}[c]{@{}l@{}}$\beta$=91.15$^{\rm o}$   \\ $\beta$=90.86$^{\rm o}$\end{tabular}                                                         \\
\cline{2-8}
		& cfg6                   & C2 (5)                         & \begin{tabular}[c]{@{}l@{}}DFT   \\ MLP\end{tabular}                      & \begin{tabular}[c]{@{}l@{}}34.780  \\ 34.792\end{tabular}                & \begin{tabular}[c]{@{}l@{}}4.081   \\ 4.085\end{tabular}                  & \begin{tabular}[c]{@{}l@{}}28.389  \\ 28.389\end{tabular}               & \begin{tabular}[c]{@{}l@{}}$\beta$=91.25$^{\rm o}$   \\ $\beta$=91.23$^{\rm o}$\end{tabular}                                                         \\
\cline{2-8}
		& cfg7                   & P21   (4)                      & \begin{tabular}[c]{@{}l@{}}DFT  \\ MLP\end{tabular}                      & \begin{tabular}[c]{@{}l@{}}20.802  \\ 20.828\end{tabular}                & \begin{tabular}[c]{@{}l@{}}4.088  \\ 4.087\end{tabular}                  & \begin{tabular}[c]{@{}l@{}}28.370   \\ 28.370\end{tabular}               & \begin{tabular}[c]{@{}l@{}}$\beta$=91.15$^{\rm o}$  \\ $\beta$=91.07$^{\rm o}$\end{tabular}                                                         \\
\cline{2-8}
		& cfg8                   & C2   (5)                       & \begin{tabular}[c]{@{}l@{}}DFT   \\ MLP\end{tabular}                      & \begin{tabular}[c]{@{}l@{}}48.482  \\ 48.475\end{tabular}                & \begin{tabular}[c]{@{}l@{}}4.088   \\ 4.091\end{tabular}                  & \begin{tabular}[c]{@{}l@{}}28.378   \\ 28.350\end{tabular}               & \begin{tabular}[c]{@{}l@{}}$\beta$=91.13$^{\rm o}$    \\ $\beta$=90.79$^{\rm o}$\end{tabular}                                                         \\
\cline{2-8}
		& cfg9                   & R3m (160)                      & \begin{tabular}[c]{@{}l@{}}DFT   \\ MLP  \\ Expt.\cite{ChemicalReviews_Tan2023} \end{tabular} & \begin{tabular}[c]{@{}l@{}}4.073  \\ 4.076\\   4.03\end{tabular}     & \begin{tabular}[c]{@{}l@{}}4.073   \\ 4.076   \\ 4.03\end{tabular}     & \begin{tabular}[c]{@{}l@{}}28.911   \\ 28.813    \\ 28.75\end{tabular} & \begin{tabular}[c]{@{}l@{}} $\gamma$=120$^{\rm o}$\\    $\gamma$=120$^{\rm o}$   \\ $\gamma$=120$^{\rm o}$\end{tabular}                                              \\
\cline{2-8}
		& cfg10                  & R$\bar{3}$m   (166)                   & \begin{tabular}[c]{@{}l@{}}DFT  \\ MLP   \\ Expt.\cite{ChemicalReviews_Tan2023} \end{tabular} & \begin{tabular}[c]{@{}l@{}}3.994  \\ 3.991   \\ 4.01\end{tabular}     & \begin{tabular}[c]{@{}l@{}}3.994    \\ 3.991   \\ 4.01\end{tabular}     & \begin{tabular}[c]{@{}l@{}}27.764   \\ 27.895   \\ 28.39\end{tabular} & \begin{tabular}[c]{@{}l@{}}$\gamma$=120$^{\rm o}$   \\ $\gamma$=120$^{\rm o}$  \\ $\gamma$=120$^{\rm o}$\end{tabular}                                              \\
\cline{2-8}
		& cfg11                  & Pc   (7)                       & \begin{tabular}[c]{@{}l@{}}DFT   \\ MLP\end{tabular}                      & \begin{tabular}[c]{@{}l@{}}7.181  \\ 7.179\end{tabular}                  & \begin{tabular}[c]{@{}l@{}}7.941   \\ 7.919\end{tabular}         & \begin{tabular}[c]{@{}l@{}}28.719  \\ 28.942\end{tabular}               & \begin{tabular}[c]{@{}l@{}}$\beta$=90.02$^{\rm o}$   \\ $\beta$=90.20$^{\rm o}$\end{tabular}                                                         \\
		\toprule[0.25mm]
		\multirow{9}{*}{1T}  & cfg12                  & P2$_1$/c   (14)                   & \begin{tabular}[c]{@{}l@{}}DFT    \\ MLP\end{tabular}                      & \begin{tabular}[c]{@{}l@{}}7.198  \\ 7.228\end{tabular}                  & \begin{tabular}[c]{@{}l@{}}7.942  \\ 7.910\end{tabular}                  & \begin{tabular}[c]{@{}l@{}}9.616   \\ 9.626\end{tabular}                 & \begin{tabular}[c]{@{}l@{}}$\beta$=93.29$^{\rm o}$   \\ $\beta$=93.34$^{\rm o}$\end{tabular}                                                         \\
\cline{2-8}
		& cfg13                  & P3m1   (156)                   & \begin{tabular}[c]{@{}l@{}}DFT   \\ MLP   \\ Expt.\cite{ChemicalReviews_Tan2023} \end{tabular} & \begin{tabular}[c]{@{}l@{}}3.988   \\ 3.989\\    4.01\end{tabular}     & \begin{tabular}[c]{@{}l@{}}3.988  \\ 3.989   \\ 4.01\end{tabular}     & \begin{tabular}[c]{@{}l@{}}9.530   \\ 9.349   \\ 9.76\end{tabular}    & \begin{tabular}[c]{@{}l@{}}$\gamma$=120$^{\rm o}$  \\ $\gamma$=120$^{\rm o}$  \\ $\gamma$=120$^{\rm o}$\end{tabular}                                              \\
\cline{2-8}
		& cfg14                  & P1   (1)                       & \begin{tabular}[c]{@{}l@{}}DFT    \\     \\      \\ MLP\end{tabular}  & \begin{tabular}[c]{@{}l@{}}24.216   \\     \\     \\ 24.207\end{tabular} & \begin{tabular}[c]{@{}l@{}}24.300\\     \\     \\    24.292\end{tabular} & \begin{tabular}[c]{@{}l@{}}9.584   \\     \\     \\ 9.580\end{tabular} & \begin{tabular}[c]{@{}l@{}} $\alpha$=88.94$^{\rm o}$   \\ $\beta$=88.99$^{\rm o}$   \\ $\gamma$=60.22$^{\rm o}$   \\ $\alpha$=88.98$^{\rm o}$   \\ $\beta$=89.02$^{\rm o}$   \\ $\gamma$=60.20$^{\rm o}$\end{tabular} \\
		\toprule[0.4mm]
	\end{tabular}
	\label{tab:TableS3}
\end{table}

\newpage
\clearpage
\bibliography{Reference}
\bibliographystyle{naturemag}